%% file: Amended_version.tex
\documentclass[nofootinbib,aps,pra,superscriptaddress,onecolumn, titlepage=false]{revtex4-2}

\input{def_new}

\begin{document}
		\title{Quantum gravity states, entanglement graphs and second-quantized tensor networks}
	\author{Eugenia Colafranceschi}
	\email{eugenia.colafranceschi@nottingham.ac.uk}
	\affiliation{School of Mathematical Sciences and Centre for the Mathematics and Theoretical Physics of Quantum Non-Equilibrium Systems, University of Nottingham, University Park Campus, Nottingham NG7 2RD, United Kingdom}

	\author{Daniele Oriti}
	\email{daniele.oriti@physik.lmu.de}
	\affiliation{Arnold Sommerfeld Center for Theoretical Physics, \\ Ludwig-Maximilians-Universit\"at München \\ Theresienstrasse 37, 80333 M\"unchen, Germany}
	\begin{abstract}
In recent years, the import of quantum information techniques in quantum gravity opened new perspectives in the study of the microscopic structure of spacetime. We contribute to such a program by establishing a precise correspondence between the quantum information formalism of tensor networks (TN), in the case of projected entangled-pair states (PEPS) generalised to a second-quantized framework, and group field theory (GFT) states, and by showing how, in this quantum gravity approach, discrete spatial manifolds arise as entanglement patterns among quanta of space, having a dual representation in terms of graphs and simplicial complexes. We devote special attention to the implementation and consequences of the label independence of the graphs/networks, corresponding to the indistinguishability of the space quanta and representing a discrete counterpart of the diffeomorphism invariance of a consistent quantum gravity formalism. We also outline a relational setting to recover distinguishability of graph/network vertices at an effective and physical level, in a partial semi-classical limit of the theory. %We first provide a basis-independent prescription to construct entanglement graphs with arbitrary connectivity, and define a scalar product which enables to compare different graphs according to combinatorial criteria. This procedure is defined for distinguishable space quanta, dual to labeled graph-vertices. As the latter do not posses any physical meaning,  we then remove them to obtain label-independent entanglement graphs, as a discrete analogue of diffeomorphism invariant states, within the Fock space of GFT states. We then outline a relational setting to recover effective distinguishability of vertices in a partial semi-classical limit of the theory. We conclude by presenting the dictionary between the GFT entanglement graphs and the TN-decomposition of projected entangled-pair states (PEPS). Within this correspondence, tensor networks acquire a natural geometric interpretation and a second-quantized, diffeomorphism invariant description.
	\end{abstract}
	\maketitle
	%\tableofcontents
	
	\section*{Introduction}
		\addcontentsline{toc}{section}{Introduction}
		Background independent approaches to the problem of quantum gravity postulate that, at the fundamental
		level, the continuous spacetime geometry dissolves into a microstructure of discrete, non-spatiotemporal entities. A central issue is then how a continuum
		spacetime emerges from the collective behaviour of the latter. A crucial role in this process seems to be played by the quantum phenomenon of entanglement. In fact several results, in quantum gravity contexts and beyond, point out a relation between entanglement and spacetime geometry and topology.  
		We mention a few of them only, out of a very large body of work.
		Within the AdS/CFT correspondence~\cite{Maldacena_1999}, the Ryu and
		Takayanagi formula~\cite{PhysRevLett.96.181602,Ryu_2006} relates the entanglement
		entropy between subsets of degrees of freedom in the boundary CFT associated to distinct regions to the area of minimal surfaces connected to that boundary regions in the dual $\text{AdS}$
		 spacetime. In~\cite{articleRaa}, it was shown that entanglement between two spacetime regions, as measured by the mutual information, is closely related to the connectivity of spacetime, as indicated by the scaling of the correlation with distance. A relation between entanglement and geometry shows up also at a dynamical level: via the holographic dictionary based on the Ryu-Takayanagi formula, the \virg{entanglement first law}~\cite{Blanco_2013} in CFT translates into  Einstein's equations linearized about pure AdS spacetime~\cite{lashkari2013gravitational,Faulkner_2014,swingle2014universality}. A different perspective on the subject is given in~\cite{PhysRevD.95.024031}, where gravity emerges from completely abstract quantum degrees of freedom: specifically, spatial geometries are constructed out of the entanglement pattern of abstract quantum states in a Hilbert
		space. This last work in fact hints at the same idea that we realize here concretely and in some detail, and in the context of a well-established quantum gravity formalism.

	The outlined scenario suggests that, in order to carry out the emergent spacetime program, we need to efficiently describe entanglement in the states of a many-body system (the collection of pre-geometric quanta) and concurrently relate such entanglement to geometric properties of a spatial or spatiotemporal structure that can be associated to the same states. In this paper we detail a correspondence between the quantum states of  the quantum gravity formalism of group field theory (GFT) \cite{oriti2011microscopic} and the quantum information language of tensor networks (TN) \cite{ran2017lecture,perezgarcia2006matrix, ORUS2014117,Bridgeman_2017}, which defines a promising framework to perform both tasks, thanks to the importing of quantum information techniques in a proper quantum gravity setting.

	A TN is a collection of tensors contracted according to a certain pattern. A single tensor is graphically represented as a node with open legs, one for each tensor index, and interpreted as a map from the degrees of freedom attached to a set of (input) legs to the complementary set of degrees of freedom on the remaining (output) legs. When only outputs are present, the tensor can thus be regarded as the state of a quantum system (say a \virg{particle}) living in the Hilbert space associated to all tensor legs. In this picture, which is the one we focus on, the tensor network describes the state of a many-body system, with a node for each particle. As the contraction of tensors - the gluing of legs - generally induces entanglement between the involved degrees of freedom, the tensor network is able to encode, in its combinatorial and quantum information, the entanglement pattern of a many-body state. As consequence, some TN satisfy entanglement area laws, a feature that makes them promising candidates for modelling holographic states in the context of the AdS/CFT correspondence, and more generally in a (quantum) gravitational context. Among the large literature exploring this application of tensor networks, we cite \cite{PhysRevD.86.065007,swingle2012constructing}, where it was showed that entanglement renormalization~\cite{PhysRevLett.99.220405} on a certain class of  many-body states gives rise to a 
network in an emergent holographic dimension and that, at a quantum critical point,  such a network reproduces a  discretized AdS space. More recently, tensor networks have proved to exhibit several other aspects of holographic duality \cite{Hayden_2016,Qi_2017,caputa2020building}, and have been used indeed to reproduce states in the AdS/CFT context \cite{Pastawski_2015,Almheiri_2015,Bao_2015,Yang_2016,Bhattacharyya_2016,Miyaji_2017,bao2019holographic,Bao_2019}.

GFT is the theory of a (bosonic) field defined on a group manifold; the excitations of the field, interpreted as quanta of space, are represented as fundamental simplices whose geometric properties are encoded in the group-theoretic variables of the field domain. By gluing the fundamental simplices to each other one can build up a discrete spatial manifold of arbitrary topology. As we are going to make explicit in the following, the gluing of different simplices is given by the entanglement between their degrees of freedom; the resulting simplicial complexes thus correspond  to \textit{entanglement patterns} among quanta of space. This is particularly evident when adopting a spin network representation: each simplex is depicted as a vertex with attached open links (i.e. links whose other ends terminate at univalent vertices); the links are dual to the faces of the simplex and carry the group variables describing its geometry. In such a dual picture, the entanglement distribution encoded in a generic GFT state is associated to a network (or, in a quantum gravity language, to a \textit{graph}). The interaction processes of the field quanta, governed by a quantum dynamics whose elementary blocks are determined by the non-local interaction kernel of GFT action, result in the combination of the simplices (the GFT quanta) into higher-dimensional simplicial complexes, to be understood as discrete counterpart of spacetime manifolds. The perturbative expansion of the GFT partition function thus returns a sum over such complexes, with Feynman amplitudes being discrete gravity path integrals on the simplicial lattices bounded by the gluings of simplices encoded in the GFT quantum states; or, equivalently, spin foam models expressing the evolution of the dual spin network states. In that sum we can find a combination of strategies for lattice quantum gravity: quantum Regge calculus \cite{Hamber2009QuantumGO} and dynamical triangulations \cite{Ambjorn:2012jv}, and the spinfoam amplitudes \cite{Perez_2003}, for the spin network states of loop quantum gravity (LQG) \cite{Thiemann2007}. In this perspective, GFT can be seen as second-quantized many-body reformulation of LQG \cite{Oriti_2016LQG}, in addition of being a direct group-theoretic enrichment of random tensor models \cite{Rivasseau_2016,Gurau_2012} (their common framework being often referred to as \virg{tensorial group field theories}).

Therefore, both formalisms of group field theory and tensor networks rely on graphical structures built up from entanglement. In this work, we make this shared feature explicit and more precise, at the same time strengthening and generalizing the correspondence between quantum gravity states and tensor networks, building on previous work which had already pointed out the relation between the LQG spin networks and particular TN decompositions, for example \cite{PhysRevB.86.195114,Singh_2010}, and carried over a first-quantized version of the GFT framework in \cite{Chirco_2018}.  On the one hand, our work advances the description of discrete geometries in GFT and the understanding of the parallel relation between entanglement and geometry; on the other, it enriches the TN language with insights from quantum gravity, overall defining a precise mathematical setting to merge tensor network techniques with that of a background-independent quantum gravity formalism. The benefits of the improved correspondence go both ways.

From the perspective of GFT quantum gravity, we provide a rigorous mathematical formulation of the presence, in the GFT Hilbert space, of states associated to arbitrarily connected graphs, clarifying also the entanglement origin of the latter. From the perspective of tensor network applications to quantum gravity (for example in the AdS/CFT context) this also enable us to show how the idea of entanglement generating geometry (area and volume realizations) and topology (the combinatorial structure of graphs), can be made concrete and explicit, at least in a discrete geometric context, based on (and already suggested by) results in LQG, spin foam models and GFT itself. This is an immediate improvement over existing applications of tensor network ideas in this direction. Indeed, while an interpretation of networks/graphs as discrete geometries is already present in the TN context, this is (severely) limited to simplicial complexes considered only in their combinatorial structure, with a geometric interpretation following from using the graph distance as metric. GFT graphs, on the other hand, carry additional quantum geometric degrees of freedom, and it is this additional structure that allows a richer entanglement/geometry correspondence, as we are going to show. 

As far as the correspondence between tensor networks and spin network graphs is concerned, the main novelty of our work respect to the aforementioned literature, and other applications of random tensor network ideas \cite{Qi_2017}, is the generalization of such correspondence to a second quantized setting, which provides the TN framework with a Fock space structure and concurrently carries a strong physical implication: the attainment of (a discrete version of) diffeomorphism invariance for the structures involved. Let us expand on this key point. When moving to the proper second-quantized formulation of GFT, a crucial difference with the quantum information language arises:  while the GFT quanta are indistinguishable, the nodes of a tensor network, as normally defined, are not. In a quantum gravity model, the indistinguishability of the building blocks of space is a necessary condition for background independence. In fact, it can be understood as a discrete counterpart of invariance under diffeomorphisms, as vertex labels play the role analogous to \virg{coordinates} over an abstract combinatorial pattern. This is why our work offers a further improvement compared to existing construction based on random tensor networks in \cite{Qi_2017}: without a suitable invariance under relabelling of the random tensor networks, their interpretation as geometries is probably incomplete. The possible way out would be to give the labels associated to tensor network nodes some physical characterization, and this is indeed another point for which we illustrate a suitable concrete realization.

In this paper, in fact, we show that GFT entanglement graphs are in fact generalised (and second-quantized) tensor networks that, in addition to having a direct simplicial-geometry interpretation, naturally satisfy (a discrete version of) invariance under relabelling/diffeomorphisms, as a consequence of the bosonic statistics of their nodes/vertices. In particular, GFT entanglement graphs can be seen as generalised random tensor networks, whose probability distribution is determined by the GFT model governing their dynamics.
%From another perspective, we can say that, within the GFT framework, tensor networks are promoted to a proper quantum gravity formalism, as they acquire a natural geometric/gravitational interpretation and a second-quantized, diffeomorphism invariance description. 
We also show how distinguishability of vertices can be recovered at a relational and effective level, by coupling the GFT field to an additional degree of freedom playing the role of a physical reference frame, in the spirit of the relational strategy typically employed in the quantum gravity context to define physical (thus, diffeomorphism invariant) observables in absence of preferred notions of space, time and locality.

Let us finally remark that, beside its quantum information aspects, our work, clarifying the way graph structures are encoded in the GFT formalism, and how the usual spin network states associated to connected graphs are embedded in the Fock space of GFT (and thus, spin foam) models, will be a strong basis also for the definition and analysis of combinatorially non-local observables (such as curvature operators) in this quantum gravity formalism.

Our work is organized as follows. After introducing, in Section \ref{GFT}, the GFT formalism, we present in Section \ref{graph} the graph theory tools we will use throughout the paper: the encoding of combinatorial patterns into matrices and the related notions of labelled- and unlabelled-graphs, i.e. graphs made of distinguishable and indistinguishable vertices, respectively. 
We then outline how to construct GFT states associated to graphs with arbitrary connectivity: we first provide, in Section \ref{label}, a basis-independent prescription to define states associated to labelled-graph, working in the pre-Fock space of the theory; in Section \ref{unlabel} we then implement vertex-relabelling invariance, obtaining states of unlabelled-graphs in the GFT Fock space. In Section \ref{csp} we define a scalar product which compares graph states independently of the vertex-labelling, with the criterion of maximising the overlap between their combinatorial structures. We conclude the analysis on graph states with Section \ref{dist}, where we show how an effective and relational notion of distinguishability of vertices can be recovered by adding new degrees of freedom (with the interpretation of discrete matter fields) to the GFT model. We then introduce, in Section \ref{TN}, the TN formalism and finally present, in Section \ref{TN}, the dictionary between group field theory states and tensor networks, explaining  how GFT (labelled- and unlabelled-)graph states can be read as precise classes of tensor network states (PEPS).

\section{The GFT formalism}\label{GFT}

A GFT is a field theory whose domain is given by ($d$ copies of) a group manifold and characterized by combinatorially non-local interactions. Let us illustrate these points. The GFT field $\phi$ is defined as follows:
\begin{equation*}\begin{split}
\phi~:&\quad G^d\qquad\rightarrow\qquad\mathbb{C}\\&\text{g}_1,...,\text{g}_d \hspace{0.8cm} \phi(\text{g}_1,...,\text{g}_d)
\end{split} 
\end{equation*}
\begin{figure}[t]
\includegraphics[width=0.5\linewidth]{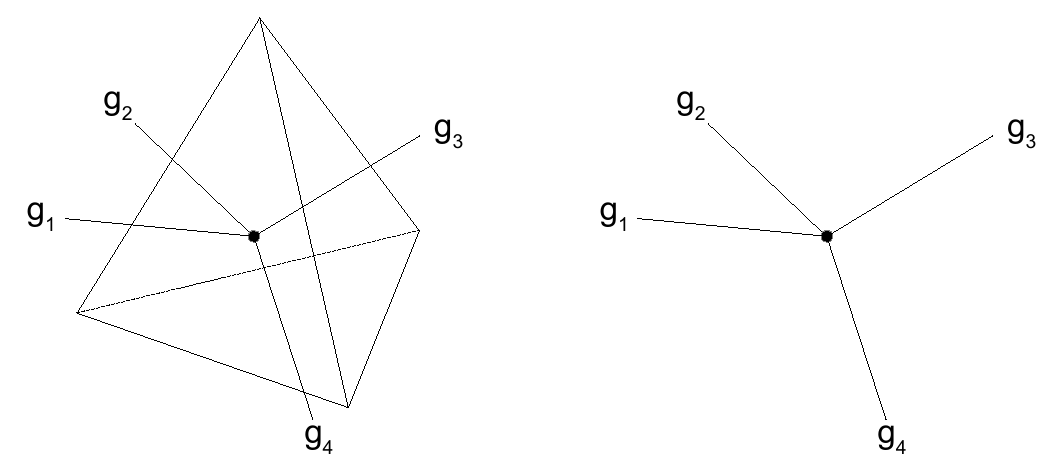}\caption{Excitation of the field $\phi(\text{g}_1,...,\text{g}_d)\in L^2(G^d/G)$ for the case $d=4$. On the left, the simplicial representation as a  tetrahedron; on the right, the dual representation as a spin network vertex.}
\label{fig:1}
\end{figure}
An excitation of the field is interpreted as a $(d-1)$-simplex, with the group variables $\text{g}_1,...,\text{g}_d$ (together with their conjugate ones under various group-theoretic Fourier transforms) associated to its faces and encoding its geometric properties. In order for $\phi(\text{g}_1,...,\text{g}_d)$ to properly describe the geometry of a $(d-1)$-simplex, it must satisfy the closure condition $\phi(h\text{g}_1,...,h\text{g}_d )=\phi(\text{g}_1,...,\text{g}_d)$, i.e. be invariant under the specified gauge transformation. Additional conditions are imposed, normally at the level of the GFT dynamics, in 4d gravitational models, where the group is taken to be $SU(2)$ or $SL(2,\mathbb{C})$ or $Spin(4)$, to ensure the proper geometric interpretation of the GFT quanta and the discrete structures they form. These geometric aspects, while of course crucial for the interpretation of the formalism in a quantum gravity context, are not directly relevant for our present purposes. As showed in Figure \ref{fig:1}, the fundamental simplex can also be represented as a  $d$-valent vertex, with an open line corresponding to each face; the latter is identified by a colour $i$,  with $i=1,...,d$, and carries the group variable $g_i$. Such a dual representation in terms of open vertices decorated by group variables makes explicit the correspondence of these quanta with the spin networks of loop quantum gravity \cite{Thiemann2007}. 

The GFT action for quantum gravity models in a simplicial context takes the general form

\begin{equation}
S_d(\phi)=\int \diff \textbf{g} \diff \textbf{g}'\phi(\textbf{g})\mathcal{K}(\{\text{g}_i\text{g}_i'^{-1}\})\phi(\textbf{g'}) + \frac{\lambda}{d+1}\int \prod_{i=1}^{d+1}\diff \textbf{g}^i\prod_{i\neq j=1}^{d+1}\mathcal{V}(\text{g}^i_j (\text{g}'^{j}_i)^{-1})\phi(\textbf{g}^1)...\phi(\textbf{g}^{d+1}),
\end{equation}
 where the bold notation $\textbf{g}$ refers to a collection of $d$ group elements, $\textbf{g}=\text{g}_1,...,\text{g}_d$, and $\diff \textbf{g}\coloneqq \diff \text{g}_1...\diff \text{g}_d$; $\mathcal{K}$ and $\mathcal{V}$ are the kinetic and the interaction kernel, respectively.
The non-local pairing of field arguments in the interaction-term determines the gluing of the fundamental $(d-1)$-simplices (in the dual picture, the linking of the corresponding vertices) into $d$-cells (graphs made of $d$-valent vertices). The interaction processes of the space quanta thus generate $d$-complexes of arbitrary topology, which are interpreted as discrete substratum of the continuum spacetime that should emerge from them in some appropriate limit. In the perturbative expansion of the GFT partition function, they are dual to the theory's Feynman diagrams. The Feynman amplitudes, on the other hand, reproduce discrete gravity path integrals, and the entire expansion can be seen as the result of merging the strategy of quantum Regge calculus \cite{Hamber2009QuantumGO} (sum
over discrete geometric data attached to a lattice) with that of dynamical
triangulations \cite{Ambjorn:2012jv} (for given geometric data, sum over all possible lattices). The Feynman amplitudes also coincide with spinfoam amplitudes \cite{Perez_2003}, as evident when expressing group functions in a group representation basis (see Section \ref{dec}). Moreover, as mentioned before, the boundary states of the $d$-complexes dual to the Feynman diagrams correspond to spin networks, the LQG candidates for the fundamental degrees of freedom of quantum geometry. Let us finally mention that GFT can be regarded as a generalization of random tensor models \cite{Rivasseau_2016,Gurau_2012}, where the combinatorial structures of the latter are enriched with group-theoretic data. As we will clarify in the following, these additional data are responsible for the characterization of the graphs associated to GFT states as patterns of entanglement among quanta.

 The GFT ladder operators satisfy bosonic commutation relations:
\begin{equation}\label{fockcomm}
[\phi(\textbf{g}^x), \phi^\dagger(\textbf{g}^y)]=\int \diff h \prod_{i=1}^d \delta (h \text{g}^x_i\text{g}_i^{y-1}),
\end{equation} 
 where the r.h.s. is the gauge invariant Dirac delta distribution on $G^d$.
The GFT Fock space is constructed starting from a vacuum state $\ket{0}$ annihilated by $\phi(\textbf{g}^x)$, with the fundamental simplices
created by the action of $\phi^\dagger(\textbf{g}^x)$ on $\ket{0}$. \medskip

\subsection{From the single-vertex Hilbert space to the Fock space}\label{preF}
As we will make extensive use of the GFT formalism in its first-quantized form, we present here the derivation of the GFT Fock space from the single-vertex Hilbert space, via the construction of a pre-Fock space. \\
A GFT vertex of valence $d$ is associated to the Hilbert space $\h= L^2(G^d/G)$, as follows by the definition of the field and the closure condition. 
%In particular, a single-vertex state  takes the form\begin{equation}\label{one}\ket{f}= \int \prod_i\diff \text{g}_i f(\text{g}_1, ..., \text{g}_d)\otimes_i \ket{\text{g}_i},\end{equation}where $\ket{\text{g}_i}$ provides a basis for the Hilbert space associated to the open link of colour $i$, $\h_i \simeq  L^2(G)$, and the wave-function $f$ satisfies $f(h\textbf{g})=f(\textbf{g})$. \medskip
Starting from that, we can consider the Hilbert space associated to a set of $V$ (distinguishable) vertices: $\h_V\coloneqq \h^{1}\otimes... \otimes \h^{V}$, where upper indices refer to vertex labels. A generic \virg{$V$-particle} state thus takes the form
\begin{equation}\label{multi}
\ket{\psi}=\int \prod_x \diff \textbf{g}^x \psi(\textbf{g}^1,...,\textbf{g}^V)\otimes_x \ket{\textbf{g}^x},
\end{equation}
where $\textbf{g}^x=\text{g}^x_1,...,\text{g}^x_d$, and $\ket{\textbf{g}^x}$ provides a basis for the single-vertex Hilbert space $\h^x$.
By taking the direct sum of the Hilbert spaces associated to all possible number of vertices $V$,  we obtain the GFT pre-Fock space:
\begin{equation}
\text{pre-}\mathcal{F}(\h)=\oplus_{V=1}^\infty  \h_V
\end{equation}
By symmetrizing each $\h_V= \h^{1}\otimes... \otimes \h^{V}$ over the vertex labels, we then obtain the Fock space of the theory:
\begin{equation}
\mathcal{F}(\h)=\oplus_{V=1}^\infty \text{sym} \left(\h^1\otimes...\otimes \h^V\right).
\end{equation}
\subsection{Spin-representation of the GFT wavefunctions} \label{dec}

%\begin{figure}[t]\includegraphics[width=0.4\linewidth]{spin.pdf}\caption{Spin decomposition of the quantum of space (a triangle) for the GFT model with $d=3$, corresponding to an excitation of the field $\phi(\text{g}_1,\text{g}_2,\text{g}_3)$. The group variables $\text{g}_i$ associated to the edges of the triangle (the open links of the dual spin network vertex) translate into the triplets given by the spin $j_i$ labelling the irreducible representations of $G$, and the pairs of basis indices $m_i,n_i$ in the vector space carrying the representation $j_i$. }\label{fig:2}\end{figure}
A function $f\in L^2(G)$, where $G$ is a compact group, can be decomposed in terms of irreducible representations of $G$ as follows (Peter-Weyl decomposition):
\begin{equation}
f(g)=\sum_{jmn} d_{j} f^{j}_{mn} D^{j}_{m n}(g),
\end{equation}
where $j\in \mathbb{N}/2$ is the \textit{spin} labelling the irreducible representations
of $G=SU(2)$; the indices $m,n$ refer to a basis in the vector space carrying the representation $j$; $d_{j}\coloneqq2j+1$ is the dimension of the latter and $D^{j}_{m n}(g)$ is the matrix representing the group element $g$. \\
%The relation between the basis $\ket{g}$ and its Peter-Weyl decomposition $\ket{j;m,n}$ is  \begin{equation}\ket{g}=\sum  \sqrt{d_{j}}\overline{D^{j}_{m n}(g)}\ket{j;m,n}\end{equation}
Starting from this, one can consider the spin decomposition of single-vertex wavefunctions $f\in L^2(G^d/G)$; for $G=SU(2)$ (which is the usual choice for the gauge group of gravity), 
\begin{equation}
f(\textbf{g})=\sum_{\textbf{j}\textbf{n}\iota} f^{\textbf{j}\iota}_{\textbf{n}} ~ \psi_{\textbf{j}\textbf{n}\iota}(\textbf{g}),
\end{equation}
where $\textbf{j}=j_1,...,j_d$ and $\textbf{n}=n_1,...,n_d$ are spins and magnetic numbers associated to the open links of the vertex, respectively, and $\iota$ is the \textit{intertwiner index} deriving from the gauge invariance of the vertex wavefunction \cite{martin2019primer}. In particular, the basis functions $\psi_{\textbf{j}\textbf{n}\iota}(\textbf{g})$ (called \textit{spin network} functions) are given by
\begin{equation}\label{spindef}
\psi_{\textbf{j}\textbf{n}\iota}(\textbf{g})= \sum_{\textbf{m}} C^{\textbf{j}\iota}_{\textbf{m}}\prod_i \sqrt{d_{j_i}} D^{j_i}_{m_i n_i}(g_i ),
\end{equation}
where $C^{\textbf{j}\iota}_{\textbf{m}}$ is the normalized intertwiner. \\
The spin decomposition clearly applies to the field operators as well; the creation operator, for example, can be written as follows: 
\begin{equation}
\phi^\dagger(\textbf{g})=\sum_{\textbf{j}\textbf{n}\iota} \phi^{\dagger\textbf{j}\iota}_{\textbf{n}} ~ \psi_{\textbf{j}\textbf{n}\iota}(\textbf{g}).
\end{equation} 
Note that $\phi^{\dagger\textbf{j}\iota}_{\textbf{n}}$ is the operator generating the spin-network basis: $\bra{\textbf{g}}\phi^{\dagger\textbf{j}\iota}_{\textbf{n}}\ket{0} =\psi_{\textbf{j}\textbf{n}\iota}(\textbf{g})$.

For more details on these group-theoretic aspects of the group field theory formalism, and of the related spin foam models and loop quantum gravity, we refer to the literature (for example, see \cite{martin2019primer}).

\section{Graphs and their adjacency matrix description}\label{graph}

In this section we introduce the graph theory notions that we will use to differentiate between combinatorial patterns implemented on distinguishable and undistinguishable quanta (and thus to proper define the GFT entanglement graphs in first- and second-quantization, respectively). For these and other notions of graph theory, we refer to \cite{ESSAM:1970zz}.

\begin{definition}[Labelled graph]
	A labelled graph $\gamma$ is an ordered set of vertices connected according to a certain pattern.
\end{definition}
 We refer to the number of vertices in a graph as its \textit{size}. A labelled graph of size $V$ can be described by a $V \times V$ matrix $A$, called \textit{adjacency matrix}, whose entries encode the adjacency relations among vertices: $A_{xy}$ takes value $1$ if vertex $x$ is connected to vertex $y$, and $0$ otherwise. Since $A$ encodes all information about $\gamma$, we refer to a graph by using both notations, i.e. $\gamma=A$.	\\	
Two graphs which differ only for the labelling of their vertices are said to be \textit{isomorphic}. Formally, two labelled graphs $\gamma=A$ and $\gamma'=A'$ of size $V$ are isomorphic if there exist a permutation $\pi$ on $V$ elements such that $A' = P_\pi A P_\pi^{-1}$, where $P_\pi$ is the matrix obtained by permuting the columns of the identity matrix.

Given an adjacency matrix $A$, we denote by $[A]$ the equivalence class of matrices obtained by permuting rows and columns of $A$:
\begin{equation}
[A] = \{A' | A' = P_\pi A P_\pi^{-1}, ~\pi\in S_V\},
\end{equation}
where $S_V$ is the set of possible permutations on $V$ elements. Note that two isomorphic graphs belongs to the same equivalence class of adjacency matrices.

\begin{definition}[Unlabelled graph]
	We define an unlabelled graph $\Gamma$ as the combinatorial pattern represented by $[A]$.  
\end{definition}

Two unlabelled graphs $\Gamma$ and $\Gamma'$ are said to be isomorphic if and only if they have a common adjacency
matrix. Moreover, two isomorphic graphs have exactly the same set of adjacency matrices.

We are interested in graphs constructed out of vertices having the same valence $d$, whose open edges are identified by colours $1,...,d$; moreover, we assume that two vertices can be connected only trough edges of the same colour. To describe these structures, we introduce generalised adjacency matrices having elements
\begin{equation}
A_{x+i,y+j}=\begin{cases}
a^i_{xy} \qquad\hfill i=j\\0 \qquad \hfill i\neq j
\end{cases}
\end{equation}
%\begin{equation}	A=	\begin{bmatrix}	0       & \textbf{a}_{12}  & \dots & \textbf{a}_{1V} \\	\textbf{a}_{21}       \\\vdots 	\\ 	\textbf{a}_{V1}       & 	\end{bmatrix} 	\end{equation}	where $	\textbf{a}_{xy}$ is a diagonal $d\times d$ matrix, 
where $a^i_{xy}=1$ if the vertices $x$ and $y$ are connected by a link of colour $i$, and $a^i_{xy}=0$ otherwise. Note that having additional data with respect to the vertex labels (i.e. the edge colours) increases the size of the adjacency matrix encoding the graph, as expected. In such a framework equivalence classes of matrices are still defined with respect to the relabelling of vertices, and the notion of unlabelled graphs naturally follows.

\section{GFT labelled-graph states}\label{label}
In GFT, gluing the \virg{quanta of space} given by excitations of the field gives rise to (discretized) spatial manifolds. Such a simplicial construction corresponds, in the dual picture, to a graph structure. As the gluing is defined by an entanglement relation, we refer to these structures as entanglement graphs. In this section we outline a prescription to construct \textit{labelled} entanglement graphs in the pre-Fock space of the theory, where the \virg{quanta of space}/vertices are \textit{distinguishable}. This is preparatory to the next section, in which we provide a definition of entanglement graphs in the truly physical (although still kinematical, i.e. before taking into account the quantum dynamics of any specific model) space of the theory, the Fock-space, where the vertices are \textit{indistinguishable} and the graphs, therefore, \textit{unlabelled}.

\subsection{Quantum geometry states associated to graphs: $\h_{\gamma}$}
Following the interpretation of graph structures in terms of discretized space, to each labelled graph $\gamma$ it is possible to associate an Hilbert space of states of quantum geometry, $\h_\gamma$, whose elements are functions of the variables associated to the links of the graph. The kinematical Hilbert space of loop quantum gravity is also constructed out of these graph-based Hilbert spaces, for all possible graphs and modulo some equivalence relations (imposing cylindrical consistency conditions)\footnote{For a discussion on the differences and similarities between the GFT and LQG Hilbert spaces, see \cite{oriti2014group}.}, locally gauge invariant; in particular, it can be realized as $\h_\gamma = L^2(G^L/G^V)$, where $L$ is the number of links and $V$ the number of vertices in $\gamma$. Note that the definition of $\h_\gamma$ is \textit{a priori} independent from regarding the graph as the result of gluing open vertices. However, by exploiting such point of view it is possible to embed labelled-graph states $\Psi_\gamma(\{g_\ell\})\in \h_{\gamma}$ into the Hilbert space $\h_V$ associated to a set of open vertices, introduced in Section \ref{preF}. The embedding of $\h_{\gamma}$ in $\h_V$ has been studied in \cite{Oriti_2016LQG}. We generalise that analysis and show how to construct in $\h_V$ graph states with arbitrary combinatorial pattern $\gamma$, in a basis-independent way.

\subsubsection{Embedding $\h_{\gamma}$ into $\h_V$} 
\begin{figure}[t]\includegraphics[width=\linewidth]{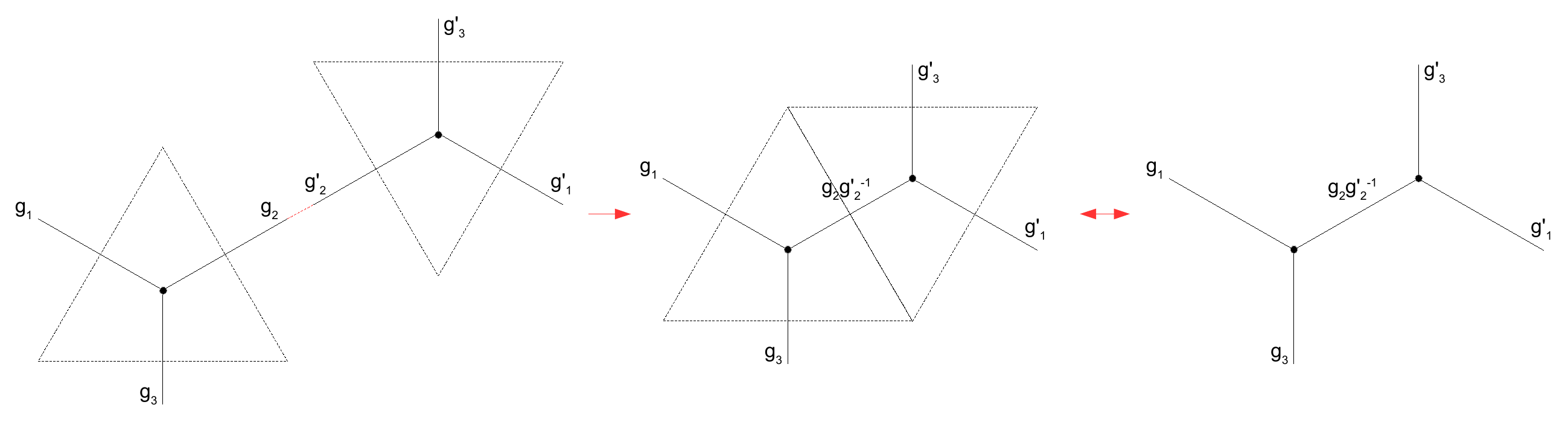}\caption{Example of the gluing of two \virg{quanta of space} of the GFT model with $d=3$, which are triangles dual to trivalent vertices. The quanta are glued along the face/open-link of colour $2$. The resulting link carries the group variable $\text{g}_2\text{g'}_2^{-1}$ given by the product of the original open-link variables $\text{g}_2$ and $\text{g'}_2$.}\label{fig:2}
\end{figure}
Consider the multi-particle state of Eq.~\eqref{multi}, which generally describes a state for a set of $V$ \textit{open} vertices. Starting from it, we can construct a special class of states in which the $V$ vertices are \textit{connected} according to a certain pattern, thanks to the entanglement among their degrees of freedom. We restrict the analysis to the case in which the connection can be realized only between edges of the same colour $i$\footnote{This restriction leads to d-colored graphs, as extensively studied in the random tensor models literature \cite{Gurau_2012,Rivasseau_2016}.}. By assigning an orientation to the edges, with the group element $g_i$ associated to the outgoing direction, the gluing of two vertices shows up as follows: the vertices $x$ and $y$ are connected along a link of colour $i$ if the multi-particle wave-function $\psi(\textbf{g}^1,...,\textbf{g}^V)$ depends on the elements $g^x_i$ and $g^y_i$ only through the product $g^x_i g_i^{y-1}$. The two vertices are then said to form an internal link $\ell=(x,y;i)$  (where $i$ is the colour of the link, $x$ and $y$ are the source and target vertices, respectively) which carries the group element $g_\ell=g^x_i g_i^{y-1}$ (see Figure \ref{fig:2}). Starting from the generic multi-particle wave-function of Eq.~\eqref{multi}, such gluing is realized by averaging through the right action of the group on the two open links carrying $g^x_i$ and $g^y_i$: 
\begin{equation}\begin{split}\label{conv}
\int \diff h\psi(...,g^x_i h, ..., g^y_i h,...)=	\psi(...,g^x_i g_i^{y-1},...).
\end{split}
\end{equation}
In fact, the convolution on the group element $h$ forces the wave-function $\psi$ to depend on $g^x_i$ and $g^y_i$ through the product $g^{x}_ig^{y-1}_i$ representing the group variable associated to the internal link $\ell=(x,y;i)$. By using the gluing prescription defined in Eq.~\eqref{conv}, a generic labelled-graph wave-function $\Psi_{\gamma}(\{g_\ell\})\in \h_{\gamma}$ can thus be seen as the result of gluing the arguments of a multi-particle wave-function $\psi(\{\textbf{g}^x\})\in\h_V$ according to the combinatorial pattern of $\gamma$: \begin{equation}\begin{split}\label{emb}
\Psi_\gamma(\{g_\ell=g^{x}_ig^{y-1}_i\})=\int \left(\prod_{\ell\in \gamma} \diff h_\ell\right) ~ \psi(...,g^x_ih_\ell, ...,g_i^{y}h_\ell,...),
\end{split}
\end{equation}
where $\ell=(x,y;i)$ refers to the links of $\gamma$. Eq.~\eqref{emb} represents the embedding of $\Psi_{\gamma}(\{g_\ell\})\in \h_{\gamma}$ in $\h_V$, and shows that the Hilbert space $\h_V$ contains, among its elements, states associate to the labelled graph $\gamma$. This result was presented in \cite{Oriti_2016LQG} as part of a broader analysis on the possibility to regard group field theory as a second quantization of loop quantum gravity. In \cite{Oriti_2016LQG} it is also shown that the scalar product on $\h_\gamma$ is the one induced by $\h^V$, therefore $\h_\gamma$ is an Hilbert subspace of $\h_V$: $\h_{\gamma} \subset \h^V$. Moreover, this induced Hilbert space coincides with the one used, for the same graph, in canonical loop quantum gravity. 

\subsection{Constructing graph states with arbitrary combinatorial pattern in $\h_V$}\label{linkmap}

We generalise and make more detailed the result of \cite{Oriti_2016LQG} by providing a prescription to construct, in $\h_V$, labelled-graph states with arbitrary combinatorial pattern, exploiting the adjacency-matrix description of graphs. We start by defining a class of  operators $\mathds{P}_i^{x\otimes y}$, called \textit{link maps}, which glue different vertices by projecting open-link states of the same colour into the internal-link subspace, i.e. the subspace invariant under the right action of the group on the open links to be glued: $\text{Inv}_R(\h^{x}_{i} \otimes \h^{y}_{i})$.
\begin{definition}[Link map]
	The gluing of two vertices  $x$ and $y$ along their open links of colour $i$ is performed by the map $\mathds{P}_i^{x\otimes y}:\h^{x}_i \otimes \h^{y}_i  ~\rightarrow ~ \text{Inv}_R(\h^{x}_{i} \otimes \h^{y}_{i})$ defined as follows:
	\begin{equation}
	\mathds{P}_i^{x\otimes y}\coloneqq\int \diff \text{h}_i^{xy}\diff g^x_i\diff g^y_i ~ ~\kbc{g^x_i }{g^x_i \text{h}_i^{xy}}{}\otimes \kbc{g^y_i }{g^y_i \text{h}_i^{xy}}{}
	\end{equation}
	where $\text{h}_i^{xy}=\text{h}_i^{yx}$.
\end{definition} 
\noindent When acting on a multi-particle state $\ket{\psi}\in \h_V$, the link map $\mathds{P}_i^{x\otimes y}$ realizes the convolution of Eq.~\eqref{conv}:
\begin{equation}
\mathds{P}_i^{x\otimes y}\ket{\psi}=\int \prod_x \diff \textbf{g}^x  \int \left(\prod_{\ell\in \gamma} \diff h_\ell\right) ~ \psi(...,g^x_ih_\ell, ...,g_i^{y}h_\ell,...)\otimes_x \ket{\textbf{g}^x}.
\end{equation}
We can then construct a graph state with arbitrary combinatorial structure $\gamma$ by applying to a multi-particle state $\ket{\psi}$ the link maps according to the adjacency matrix $A$ of $\gamma$:
\begin{equation}\begin{split}
\ket{\psi_\gamma}=& \prod_{x < y}\prod_{i: a^i_{xy}=1}\mathds{P}_i^{x\otimes y} 	\ket{\psi}\\=&\int \prod_{x} \diff \textbf{g}^x  \int \prod_{x<t_i(x)}\diff \textbf{h}^{x\textbf{t}(x)}\psi(\{g^x_i h^{xt_i(x)}_i\}) \otimes_x	\ket{\textbf{g}^x },
\end{split}
\end{equation}
with $\diff \textbf{h}^{x\textbf{t}(x)}\coloneqq \diff \text{h}_1^{x\text{t}_1(x)}...\diff \text{h}_d^{x\text{t}_d(x)}$, where $t_i(x)$ is a tensor encoding the combinatorial pattern of the graph: $t_i(x)=y$ if $a^{xy}_i=1$, and $t_i(x)=0$ if $a^{xy}_i=0$; the gluing elements $h^{xt_i(x)}_i$ are such that $h^{xy}_i= h^{yx}_i$, and $h_i^{x0}=e$ (where $e$ is the identity element). The wave-function of the resulting state $\ket{\psi_\gamma}$ is thus associated to a graph with internal links $\ell=(x,t_i(x);i)$:
\begin{equation}\begin{split}
\langle \textbf{g}^1,...,\textbf{g}^V|\psi_\gamma\rangle = \int\prod_{x<t_i(x)} \diff \textbf{h}^{x\textbf{t}(x)}\psi(\{g^x_i h^{xt_i(x)}_i\})=\Psi_\gamma(\{g_\ell=g^{x}_ig^{t_i(x)-1}_i\})
\end{split}
\end{equation}
Note that, since the gluing operation is a projection from $\h_V$ to $\h_{\gamma}\subset \h_V$, given a graph state in $\h_\gamma$, a corresponding multi-particle state in $\h_V$ for the pre-gluing phase is not uniquely defined. Let us finally remark that the provided prescription is basis-independent, as it is defined through the action of projection operators on the Hilbert spaces associated to the vertices. 
In the spin representation, the link maps are in fact a combination of Kronecker deltas identifying spin labels and vector labels in the corresponding representation spaces, for the open links that are being glued. This will be made explicit and exploited in the next section.

\subsubsection{Comparing graph states of equal size}
As mentioned before, in \cite{Oriti_2016LQG} it was showed that the scalar product on $\h_\gamma$ is the one induced by $\h^V$. 
Following our generalised construction of graph states in $\h_V$, we show that the scalar product on $\h_V$ allows also to compare graph states with the same number $V$ of vertices but possibly different combinatorial structure. Note that, two quantum states associated to graphs with different number of vertices are necessarily orthogonal, due to the structure of the GFT (pre-)Fock space.\\
We work in the spin network representation, and thus start by presenting the expansion of a graph wave-function embedded in $\h_V$ in such a basis. 
Bold symbols refers to collections of edge variables, e.g. $\textbf{j}^x=j^x_1...j^x_d$, while a vector notation is used for sets of vertex variables, e.g. $\vec{\textbf{j}}=\textbf{j}^1...\textbf{j}^V$. We also recall that $\ell=(x,y;i)$ is the link of colour $i$ connecting $x$ (source vertex) to $y$ (target vertex) and carrying the variable $g_\ell=g^x_i g_i^{y-1}$. The graph wave-function $\Psi_\gamma$ embedded in $\h_V$ takes, in spin representation, the following form:
\begin{equation}\begin{split}
\Psi_\gamma(\{g_\ell\})=&
\int \prod_{x< y} \diff \text{h}_i^{x y} \psi(\{g^x_i h^{xy}_i\})=\int \prod_{x < y} \diff \text{h}_i^{x y}\sum_{\vec{\textbf{j}}\vec{\iota}} \psi^{\vec{\textbf{j}}}_{\vec{\textbf{p}}\vec{\textbf{n}}}~ C^{\vec{\textbf{j}}\vec{\iota}}_{\vec{\textbf{p}}} C^{\vec{\textbf{j}}\vec{\iota}}_{\vec{\textbf{m}}} ~\prod_{x,i}d_{j^x_i} D^{j^x_i}_{m^x_i n^x_i}(g^x_i h^{xt_i(x)}_i)\\=&\sum \psi^{\vec{\textbf{j}}\vec{\iota}}_{\vec{\textbf{n}}} ~ C^{\vec{\textbf{j}}\vec{\iota}}_{\vec{\textbf{m}}}\int \prod_{x < y} \diff \text{h}_i^{x y}\prod_{x,i}\sqrt{d_{j^x_i}} D^{j^x_i}_{m^x_i n^x_i}(g^x_i h^{xt_i(x)}_i)
\end{split}
\end{equation}
with $C^{\vec{\textbf{j}}\vec{\iota}}_{\vec{\textbf{p}}}\coloneqq \prod_x C^{\textbf{j}^x\iota^x}_{\textbf{p}^x}$ and  
\begin{equation}
\psi^{\vec{\textbf{j}}\vec{\iota}}_{\vec{\textbf{n}}}\coloneqq\psi^{\vec{\textbf{j}}}_{\vec{\textbf{p}}\vec{\textbf{n}}}C^{\vec{\textbf{j}}\vec{\iota}}_{\vec{\textbf{p}}}\prod_{x,i}\sqrt{d_{j^x_i}},
\end{equation}
and the sum is over all repeated indices. By performing the integral over the gluing elements\footnote{We use the relation
\begin{align}
	&\int \diff h D^{j'}_{m'n'}(h)D^j_{mn}(h)=\frac{1}{d_j}\delta_{j',j}\delta_{m',m}\delta_{n',n},
\end{align}where $d_j\coloneqq2j+1$.} we obtain
\begin{equation}\begin{split}\label{a}
\Psi_\gamma(\{g_\ell\})=\sum\Psi_{\gamma~ \{n^{x}_i\}_{\text{open}}}^{\{j_i^{xt_i(x)}\}\vec{\iota}}~C^{\{j_i^{xt_i(x)}\}\vec{\iota}}_{\vec{\textbf{m}}}\prod_{x,i:x<t_i(x)} \sqrt{d_{j^{xt_i(x)}_i}} D^{j^{xt_i(x)}_i}_{m^x_i m^{t_i(x)}_i}(g^x_i g^{t_i(x) -1}_i)\prod_{x,i:t_i(x)=0}\sqrt{d_{j^x_i}} D^{j^x_i}_{m^x_i n^x_i}(g^x_i )
\end{split}
\end{equation} where
\begin{equation}
\Psi_{\gamma~ \{n^{x}_i\}_{\text{open}}}^{\{j_i^{xt_i(x)}\}\vec{\iota}}\coloneqq \Psi_{\gamma~ \{n^{x}_i\}_{\text{open}}}^{j_1^{1t_1(1)}...j_d^{Vt_d(V)}i^1...i^V}=\psi^{\vec{\textbf{j}}\vec{\iota}}_{\vec{\textbf{n}}}~\prod_{x,i:x< t_i(x)}\frac{1}{\sqrt{d_{j^{xt_i(x)}_i}}}\delta_{j^x_i,j^{xt_i(x)}_i}~\delta_{j^{t_i(x)}_i,j^{xt_i(x)}_i}~\delta_{n^x_i,n^{t_i(x)}_i}\end{equation}
with the conventions $j^{xy}_i=j^{yx}_i$ and $j^{x0}_i=j^x_i$. In Eq.~\eqref{a} we can recognize basis wave-functions for the labelled-graph state:
\begin{equation}\begin{split}
\label{b}
\theta_{\gamma \{n^x_i\}_{\text{open}}}^{\{j_i^{xt_i(x)}\}\vec{\iota}}(\{g_\ell\}) =	C^{\{j_i^{xt_i(x)}\}\vec{\iota}}_{\vec{\textbf{m}}}\prod_{x,i:x<t_i(x)}\sqrt{d_{j^{xt_i(x)}_i}} D^{j^{xt_i(x)}_i}_{m^x_i m^{t_i(x)}_i}(g^x_i g^{t_i(x) -1}_i)\prod_{x,i:t_i(x)=0}\sqrt{d_{j^x_i}} D^{j^x_i}_{m^x_i n^x_i}(g^x_i ).
\end{split}
\end{equation}
In fact, Eq.~\eqref{a} can be rewritten as
\begin{equation}\begin{split}
\Psi_\gamma(\{g_\ell\})=&\sum_{\{j_\ell\}\vec{\iota}} \Psi_{\gamma \{n_\ell\}_{\text{open}}}^{\{j_\ell\}\vec{\iota}}~\theta_{\gamma \{n_\ell\}_{\text{open}}}^{\{j_\ell\}\vec{\iota}}(\{g_\ell\})
\end{split}
\end{equation}
To show that the natural scalar product in $\h_V$ allows to compare states associated to graphs of equal size $V$ but possibly different connectivity, we can restrict the attention to the basis states $\ket{\theta_\gamma(\{j_\ell,n_\ell,\vec{\iota}\})}$. We obtain that (see Appendix \ref{scalarp} for details)
\begin{equation}\begin{split}
\langle\theta_{\gamma'}(\{j'_\ell,n'_\ell,\vec{\iota'}\})|\theta_\gamma(\{j_\ell,n_\ell,\vec{\iota}\})\rangle=&\prod_x \delta_{j_i'^{xt'_i(x)},j^{xt_i(x)}_i}\prod_{x:t_i(x)= 0, t'_i(x)\neq 0}\delta_{n^{t'_i(x)}_i,n^{x}_i}\prod_{x:t_i(x)\neq 0, t'_i(x)= 0}\delta_{n'^{x}_i,n'^{t_i(x)}_i}\\&\cdot\prod_{x:t_i(x)=t'_i(x)= 0}\delta_{n'^{x}_i,n^{x}_i}  ~\delta(\vec{\iota},\vec{\iota'})
\end{split}
\end{equation}
The above expression shows that graph states with different combinatorial structures are not necessarily orthogonal. Such feature derives from the fact that, in our framework, graphs do not underpin the definition of the kinematical Hilbert space, but arise as entanglement patterns among quanta, defined in a larger (with respect to the degrees of freedom associated to each graph) (pre-)Fock space. Let us also remark that, though given a graph wave-function $\Psi_\gamma \in \h_\gamma$ the multi-particle one $\psi \in \h_V$ of the pre-gluing phase is not uniquely defined, such an ambiguity does not affect the result of the scalar product.
\subsubsection{Labelled-graph states from individually weighted vertices}\label{iwv}
Here we consider a special class of GFT states constructed out of a set of \textit{individually weighted vertices}, namely a set where each vertex is dressed with a single-particle wavefunction $f \in \h$. The interest in these states is multiple: in addition to be the simplest generalization of condensate states used in cosmology, also encoding space-connectivity information \cite{Oriti_2015,Oriti_2016}, they have been used in \cite{PhysRevD.97.066017,PhysRevLett.116.211301} to model black hole geometries; moreover, in their first-quantized expression they were put in relation to tensor networks in \cite{Chirco_2018}.\\ For this class of states, the multi-particle state of the pre-gluing phase is factorized over the single-vertex Hilbert spaces: 
\begin{equation}
\ket{\psi^{\vec{f}}}=\otimes_{x}	\ket{f_x}_x 
\end{equation}
where $\vec{f}$ denotes the set of single-vertex wave-functions: $\vec{f}=(f_1,...,f_V)$. By applying to this state the link maps according to the combinatorial structure $\gamma=A$ we obtain
\begin{equation}\begin{split}\label{iwvs}
\ket{\psi^{\vec{f}}_{\gamma}}\coloneqq&\prod_{x < y}\prod_{i: a^i_{xy}=1}\mathds{P}_i^{x\otimes y} \otimes_{x}	\ket{f^x}_x\\=&\int \prod_{x} \diff \textbf{g}^x\int \prod_{x<t_i(x)}\diff \textbf{h}^{x\textbf{t}(x)} \prod_{x}f_x(\textbf{g}^x \textbf{h}^{x\textbf{t}(x)})\otimes_x	\ket{\textbf{g}^x }_x
\end{split}
\end{equation}
This corresponds to a certain state $\ket{\Psi_\gamma}\in \h_\gamma$, with $
\Psi_\gamma(\{q_\ell=q^{x}_iq^{y-1}_i\})=\psi^{\vec{f}}_{\gamma}(\{q^{x}_iq^{y-1}_i\})$. 
Note that, given a graph function $\Psi_\gamma(\{q_\ell\})$, it is always possible to identify a set of functions $f_1,...,f_V$ that glued according to the adjacency matrix of the graph $\gamma$ give $\Psi_\gamma(\{q_\ell\})$. 
\section{GFT unlabelled-graph states}\label{unlabel}

 As we have shown, in GFT the simplicial complexes resulting from the gluing of the fundamental simplices (the \virg{quanta of space}) are encoded in the entanglement structure of multi-particle states, and represented by graphs, whose vertices are dual to the fundamental simplices. So far we considered the GFT vertices as \textit{distinguishable}, i.e. we labelled them and worked in the pre-Fock space of the theory. However, we know that vertex labels are just an auxiliary structure, which does not possess any physical meaning. In the following, we show how to remove it from our labelled-graph states by symmetrizing over the vertex labels, thereby obtaining states associated to unlabelled graphs. This also means working in the true Hilbert space of the underlying GFT, i.e. the Fock space in which only wavefunctions symmetric under permutations of the vertex set appear.
 \paragraph*{First-quantized unlabelled graph states} Given a state $\ket{\psi_\gamma}$ associated to a graph $\gamma=A$, we turn it into a state invariant under vertex-relabelling by symmetrizing over the vertex group variables:
\begin{equation}\begin{split}\label{ugs}
\ket{\psi_{\gamma}}=\int \prod_x \diff \textbf{g}^x\psi_{\gamma}(\textbf{g}^1...\textbf{g}^V)\otimes_{x}	\ket{\textbf{g}^x}_x \quad \rightarrow \quad\ket{\psi_{\Gamma}}=\int \prod_x \diff \textbf{g}^x\psi_{\Gamma}(\textbf{g}^1...\textbf{g}^V)\otimes_{x}	\ket{\textbf{g}^x}_x
\end{split}
\end{equation}
where 
\begin{equation} \label{ugwf}
\psi_{\Gamma}(\textbf{g}^1...\textbf{g}^V)=\underset{x}{\text{sym}}\left(\psi_\gamma(\{g^{x}_ig^{t_i(x)-1}_i\})\right)=\sum_\pi \int\prod_{x<t_i(x)} \diff \textbf{h}^{x\textbf{t}(x)}\psi(\{g^{\pi(x)}_i h^{xt_i(x)}_i\}),
\end{equation}
with $\pi$ referring to a permutation over $V$ elements. In $\ket{\psi_{\Gamma}}$, the vertex degrees of freedom are still entangled according to the pattern of the original labelled-graph state, but the vertices are \textit{indistinguishable}; the state is thus associated to the unlabelled graph $\Gamma=[A]$. \\Denoting by $\mathds{P}_\pi$ the operator performing the relabelling $x\rightarrow\pi(x)$, i.e. $\bra{\textbf{g}^{1},...,\textbf{g}^{V}}\mathds{P}_\pi\ket{\psi_{\gamma}}=\psi_\gamma(\textbf{g}^{\pi(1)},...,\textbf{g}^{\pi(V)})$, we can write $\ket{\psi_{\Gamma}}$ as follows:
\begin{equation}\begin{split}
\ket{\psi_{\Gamma}}=\sum_{\pi \in S_V}\mathds{P}_\pi\ket{\psi_{\gamma}}=\mathds{P}_{\text{inv}_\pi}\ket{\psi_\gamma}
\end{split}
\end{equation}
where $\mathds{P}_{\text{inv}_\pi}=\sum_{\pi \in S_V}\mathds{P}_\pi$ is the operator projecting the labelled-graph state into the subspace invariant under vertex re-labelling. 
\paragraph*{Second-quantized unlabelled graph states}

The unlabelled graph state $\ket{\psi_{\Gamma}}$ belongs, by definition, to the Fock space $\mathcal{F}(\h)$, and can be written in the second-quantized formalism as follows:
\begin{equation}\begin{split}
\label{statefock}
\ket{\psi_{\Gamma}}=&\int \prod_{x} \diff \textbf{g}^x \psi_\gamma(\textbf{g}^1,...,\textbf{g}^V) \prod_x  \phi^\dagger(\textbf{g}^x)	 \ket{0}
\end{split}
\end{equation}
In fact, the symmetry of the wavefunction is ensured by the commutativity of the field operators, which \virg{project} $\psi_\gamma$ to the Fock space. \\
So far we constructed unlabelled-graph states starting from labelled-graph ones and implementing invariance under vertex-relabelling. This is the most natural procedure as vertex labels, despite lacking a physical interpretation, are needed to define a graph. However, we could be interested in implementing an entanglement pattern directly in the Fock space. For this purpose, given an unlabelled graph $\Gamma$ of size $V$, we introduce the following $V+V$-body operator:
\begin{equation}\begin{split}\label{O}
O_\Gamma= \int \prod_x \diff \textbf{g}^{x} \int \prod_{x<\text{t}_i(x)} \diff \textbf{h}^{x \textbf{t}(x)}\prod_x
\phi^\dagger(\textbf{g}^x \textbf{h}^{x \textbf{t}(x)})\prod_x \phi(\textbf{g}^x),
\end{split}
\end{equation}
where $t_i(x)$ is the tensor encoding the connectivity of $\Gamma$. When acting on a $V$-particle basis state, $O_\Gamma$ entangles the vertex degrees of freedom according to the pattern $\Gamma$:
\begin{equation}\begin{split}\label{ubase}
O_\Gamma \prod_x \phi(\textbf{g}^x)^\dagger \ket{0} = \int \prod_x \diff \textbf{h}^{x\textbf{t}(x)}
\phi^\dagger(\textbf{g}^x \textbf{h}^{x\textbf{t}(x)})\ket{0}
\end{split}
\end{equation}
Note that, though the operator $O_\Gamma$ generates an entanglement pattern directly in the Fock space, it is still dependent from the possibility to distinguish vertices; in fact, defining the tensor $\textbf{t}(x)$ requires assigning a vertex labelling to $\Gamma$. Note also that $O_\Gamma$ can be thought of as a second-quantized version of the link maps introduced in Section \ref{linkmap}. However, it is not a projection operator, as 
further applications of $O_\Gamma$ on the state of Eq.~\eqref{ubase} leaves the latter unchanged only if the pattern $\Gamma$ is symmetric (completely connected/disconnected unlabelled-graph). In fact we have that

\begin{equation}\begin{split}\label{app}
O^2_\Gamma \prod_x \phi(\textbf{g}^x)^\dagger \ket{0}=& \int\prod_{x<\text{t}_i(x)} \diff \textbf{h'}^{x \textbf{t}(x)} \prod_{x<\text{t}_i(x)} \diff \textbf{h}^{x \textbf{t}(x)}\sum_\pi\prod_x
\phi^\dagger(\textbf{g}^{\pi(x)} \textbf{h}^{\pi(x) \textbf{t}(\pi(x))}\textbf{h'}^{x \textbf{t}(x)}) \ket{0}\\=&\int\prod_{x<\text{t}_i(x)} \diff \textbf{h'}^{x \textbf{t}(x)} \prod_{x<\text{t}_i(x)} \diff \textbf{h}^{x \textbf{t}(x)}\sum_\pi\prod_x
\phi^\dagger(\textbf{g}^{x} \textbf{h}^{x \textbf{t}(x)}\textbf{h'}^{\pi^{-1}(x) \textbf{t}(\pi^{-1}(x))}) \ket{0},
\end{split}
\end{equation}
and, in order for the r.h.s of Eq.~\eqref{app} to be proportional to the r.h.s of Eq.~\eqref{ubase}, all links in $\Gamma$ must be glued (case $t_i(x)\neq 0 ~\forall i,x$) or open (case $t_i(x)= 0 ~\forall i,x$).

\begin{comment}
\begin{equation}\begin{split}
O^2_\Gamma \prod_x \phi(\textbf{g}^x)^\dagger \ket{0}=& \int\prod_{x<\text{t}'_i(x)} \diff \textbf{h'}^{x \textbf{t'}(x)} \prod_{x<\text{t}_i(x)} \diff \textbf{h}^{x \textbf{t}(x)}\sum_\pi\prod_x
\phi^\dagger(\textbf{g}^{\pi(x)} \textbf{h}^{\pi(x) \textbf{t}(\pi(x))}\textbf{h'}^{x \textbf{t'}(x)}) \ket{0}\\=&\int\prod_{x<\text{t}'_i(x)} \diff \textbf{h'}^{x \textbf{t'}(x)} \prod_{x<\text{t}_i(x)} \diff \textbf{h}^{x \textbf{t}(x)}\sum_\pi\prod_x
\phi^\dagger(\textbf{g}^{x} \textbf{h}^{x \textbf{t}(x)}\textbf{h'}^{\pi^{-1}(x) \textbf{t'}(\pi^{-1}(x))}) \ket{0}
\end{split}
\end{equation}
In order for the r.h.s of Eq.~\eqref{app} to be proportional to the r.h.s of Eq.~\eqref{ubase}, there must exist a set of elements $\text{H}_i^{xt_i(x)}$ such that
\begin{equation}
\text{h}^{xt_i(x)}_i\text{h'}^{\pi^{-1}(x) t'_i(\pi^{-1}(x))}_i =\text{H}^{xt_i(x)}_i \quad \forall \pi,
\end{equation}
This leads to the condition
\begin{equation}
\text{h'}^{12}_i=\text{h'}^{13}_i=...=\text{h'}^{1V}_i=\text{h'}_i
\end{equation}
which implies that, in $\Gamma$, all the links of colour $i$ are glued (case $\text{h'}_i\neq e$) or open (case $\text{h'}_i=e$).
\end{comment}
 
 \subsubsection{Unlabelled-graph states from individually weighted vertices}
 Here we introduce the unlabelled version of the graph states constructed out of individually weighted vertices, defined in Section \ref{iwv}. Consider the labelled-graph state $\ket{\psi^{\vec{f}}_\gamma}$ defined in Eq.~\eqref{iwvs}; its unlabelled counterpart is given by
 \begin{equation}\begin{split}
 \ket{\psi^{\vec{f}}_{\Gamma=[A]}}=\int \prod_x \diff \textbf{g}^x\psi^{\vec{f}}_{\Gamma=[A]}(\textbf{g}^1...\textbf{g}^V)\otimes_{x}	\ket{\textbf{g}^x}_x
 \end{split}
 \end{equation}
 where
 \begin{equation} 
 \psi^{\vec{f}}_{\Gamma=[A]}(\textbf{g}^1...\textbf{g}^V)=\underset{x}{\text{sym}}\left(\psi^{\vec{f}}_{\gamma=A}(\{g^{x}_ig^{t_i(x)-1}_i\})\right)=\sum_\pi \left(\int  \prod_{x<t_i(x)} \diff \textbf{h}^{x\textbf{t}(x)} \prod_x  f_x(\textbf{g}^{\pi(x)}\textbf{h}^{x\textbf{t}(x)})	\right)
 \end{equation}
 In this formula, the notation $\Gamma=[A]$ is used to specify that the vector $\vec{f}$ refers to the adjacency matrix $A$.
 Note that
 \begin{equation}\begin{split}
 \psi_\gamma^{\vec{f}}(\textbf{g}^{\pi(1)},...,\textbf{g}^{\pi(V)}) &=\int  \diff \text{h}^{xt_i(x)}_i \prod_x  f_x(\textbf{g}^{\pi(x)}\textbf{h}^{x\textbf{t}(x)})\\&=\int  \diff \text{h}^{xt_i(x)}_i \prod_x  f_{\pi^{-1}(x)}(\textbf{g}^{x}\textbf{h}^{\pi^{-1}(x)\textbf{t}(\pi^{-1}(x))})\\&=\int  \diff \text{h'}^{xt'_i(x)}_i \prod_x  f_{\pi^{-1}(x)}(\textbf{g}^{x}\textbf{h'}^{x\textbf{t'}(x)})\\&=\psi_{\gamma'}^{\vec{f'}}(\textbf{g}^{1},...,\textbf{g}^{V}),\end{split}
 \end{equation}
 where $f'_x \coloneqq f_{\pi^{-1}(x)}$ and $\textbf{h'}^{x\textbf{t'}(x)}\coloneqq \textbf{h}^{\pi^{-1}(x)\textbf{t}(\pi^{-1}(x))}$. 
 %In spin representation:\begin{equation}\begin{split}\psi^{\vec{f}}_\gamma(\textbf{g}^{\pi(1)},...,\textbf{g}^{\pi(V)})=&\sum_{j,i}~ \prod_x f^{\textbf{j}^{\pi(x)}i^{\pi(x)}}_{x~ \textbf{n}^{\pi(x)}} ~ C^{\vec{\textbf{j}}\vec{\iota}}_{\vec{\textbf{m}}}\int \diff \text{h}_i^{x t_i(x)}\prod_x D^{j^{\pi(x)}_i}_{m^{\pi(x)}_i n^{\pi(x)}_i}(g^{\pi(x)}_i h^{xt_i(x)})\\=&\sum_{j,i}~ \prod_x f^{\textbf{j}^{x}i^{x}}_{\pi^{-1}(x)~ \textbf{n}^{x}} ~ C^{\vec{\textbf{j}}\vec{\iota}}_{\vec{\textbf{m}}}\int \diff \text{h}_i^{x t_i(x)}\prod_x D^{j^{x}_i}_{m^{x}_i n^{x}_i}(g^{x}_i h^{\pi^{-1}(x)t_i(\pi^{-1}(x))})\\=&\sum_{j,i}~ \prod_x f'^{\textbf{j}^{x}i^{x}}_{x~ \textbf{n}^{x}} ~ C^{\vec{\textbf{j}}\vec{\iota}}_{\vec{\textbf{m}}}\int \diff \text{h}_i^{x t_i(x)}\prod_x D^{j^{x}_i}_{m^{x}_i n^{x}_i}(g^{x}_i h'^{xt'_i(x)})\end{split}\end{equation}
 That is, $\mathds{P}_\pi\ket{\psi_{\gamma}^{\vec{f}}}=\ket{\psi_{\gamma'}^{\vec{f'}}}$ where 
 $A'= P_{\pi^{-1}} A P^{-1}_{\pi^{-1}}$ and $\vec{f}'=P_{\pi^{-1}} \vec{f}=(f_{\pi^{-1}(1)}, ...,f_{\pi^{-1}(V)})$.
 \begin{comment}
 Explicitly:
 \begin{equation}\begin{split}
 \ket{\psi_{\gamma'}^{\vec{f'}}}=&\prod_{x < y}\prod_{i: a^{'i}_{xy}=1}\mathds{P}_i^{x\otimes y} \otimes_{x}	\ket{f'_x}_x\\=&\prod_{x < y}\prod_{i: a^i_{\pi^{-1}(x)\pi^{-1}(y)}=1}\mathds{P}_i^{x\otimes y} \otimes_{x}	\ket{f_{\pi^{-1}(x)}}_x \\=&\int  \diff \text{h}^{xt_i(x)}_i \prod_x  f_{\pi^{-1}(x)}(\textbf{q}^{x}\textbf{h}^{\pi^{-1}(x)\textbf{t}(\pi^{-1}(x))})\otimes_{x}	\ket{\textbf{q}^x}_x
 \end{split}
 \end{equation}
 where the second equality follows from the substitutions $f'_x=f_{\pi^{-1}(x)}$, $a^{'i}_{xy}={a}^i_{\pi^{-1}(x)\pi^{-1}(y)}$.
 \end{comment}

\section{Combinatorial scalar product}\label{csp}
An unlabelled-graph state is defined by a combinatorial pattern $[A]$ and a symmetrized wavefunction depending on the variables attached to the graph elements (vertices and links). As showed in the previous section, such a state can be thought of as built up from a set of labelled-graph states related to each other by vertex-relabelling. 
We have emphasized that quantum states associated to different graphs are not orthogonal, as to be expected since they simply correspond to different entanglement patterns of the same number of quanta. At the same time, we are interested in the possibility of comparing such states and defining a precise measure of their overlap that depends directly on the underlying combinatorial pattern.

Consider the scalar product between two unlabelled-graph states, written (in the pre-Fock space) as the result of summing over labelled-graph ones:
\begin{equation}\begin{split}\label{sprefock}
\langle \varphi_{\Gamma'}|\psi_{\Gamma}\rangle = & \langle\varphi_{\gamma'}|\mathds{P}_{\text{inv}_\pi}\mathds{P}_{\text{inv}_\pi}\ket{\psi_\gamma}=\langle\varphi_{\gamma'}|\mathds{P}_{\text{inv}_\pi}\ket{\psi_\gamma}=\sum_{\pi \in S_V}\langle\varphi_{\gamma'}|\mathds{P}_\pi\ket{\psi_\gamma}
\end{split}
\end{equation}
On the basis of this expression, we define a \virg{combinatorial scalar product} which compares labelled-graph states giving relevance to the combinatorial aspect, independently on the specific vertex-labelling. It amounts to select, among all the possible relabelled versions of the states, the ones which maximise the superposition of their combinatorial structures. Equivalently, it selects the vertex-labellings corresponding to the closest adjacency matrices. Such a scalar product can be seen as a prescription to align graphs in order to maximise their overlap, and then compute the (standard) scalar product between the corresponding wave-functions. 
\begin{definition}[Combinatorial scalar product]\label{comb1} Given two graph states $\ket{\psi_{\gamma}}$ and $\ket{\varphi_{\gamma'}}$ we define their combinatorial scalar product as follows:
	\begin{equation}\begin{split}\label{comb}
	\langle \varphi_{\gamma'}|\psi_{\gamma}\rangle_{\text{comb}}\coloneqq & \langle\varphi_{\gamma'}|\mathds{P}_{\bar{\pi}}\ket{\psi_\gamma}
	\end{split}
	\end{equation}
		where the permutation $\bar{\pi}$ is such that 
	\begin{equation}
 |\langle\varphi_{\gamma'}|\mathds{P}_{\bar{\pi}}\ket{\psi_\gamma}|=\max_{\pi\in S_\text{min}}|\langle\varphi_{\gamma'}|\mathds{P}_{\pi}\ket{\psi_\gamma}| 	
	\end{equation}
with
	\begin{equation}\label{Sd}
	S_\text{min}\coloneqq \{ \pi \in S_V :  d\left(P_{\pi^{-1}} A P^{-1}_{\pi^{-1}}, B \right) = \min_{C \in [A]} d\left(C,B\right)\}
	\end{equation}
	where $A=\gamma$ and $B=\gamma'$, and $d(\cdot,\cdot)$ is a notion of distance between matrices.
\end{definition}
\noindent
Note that $\langle \varphi_{\gamma'}|\psi_{\gamma}\rangle_{\text{comb}}=1$ if the states $\ket{\psi_{\gamma}}$ and $\ket{\varphi_{\gamma'}}$ differ only for the labelling of their vertices, as expected in a setting where such a labelling is deprived of any physical meaning.

At this point, a question naturally arises: can we provide a similar prescription in the Fock space, i.e. define a scalar product which emphasizes the combinatorial structure of the states? Note that, when considering symmetric states, all permutations of the vertex labels produce the same value on the right hand side of Eq.~\eqref{comb}. Therefore, selecting a particular alignment of vertices does not affect the result. Moreover, to define such a scalar product in the Fock space is simply not possible: aligning graphs as we have done requires vertex labels, and thus to break the symmetry which underpins the very definition of the Fock space. To clarify this point, in the following we translate the combinatorial scalar product in the second-quantized formalism. We work with unlabelled-graph states written as in Eq.~\eqref{statefock} in order to recover, when breaking the Fock space symmetry, the labelled-graph wavefunctions from which they were defined\footnote{In doing this, we make a slight abuse of notation: in Eq.~\eqref{statefock} the unlabelled-graph state $\ket{\psi_\Gamma}$ is written in terms of the labelled-graph wave-function $\psi_\gamma$, but the only readable information about the latter is its symmetrized version, namely $\psi_\Gamma$; in fact, the commutativity of the creation operators $\phi^\dagger$ hides any information content about $\psi_\gamma$ which is not symmetric under vertex relabelling. Note also that, given $\psi_\Gamma$, the choice of $\psi_\gamma$ is not unique; however, this feature is not relevant for the present purpose.}. We start by rewriting Eq.~\eqref{sprefock} in a second-quantized formalism:
\begin{equation}\begin{split}
\langle\varphi_{\Gamma'}|\psi_\Gamma\rangle=&\int \prod_{x} \diff \textbf{g}^x \diff \textbf{q}^x\varphi_{\gamma'}^*(\{\textbf{q}^x\})\psi_{\gamma}(\{\textbf{g}^x\})\bra{0}\prod_x \phi(\textbf{q}^x)\prod_x \phi^\dagger(\textbf{g}^x)\ket{0}\\=&\int \prod_{x} \diff \textbf{g}^x \diff \textbf{q}^x \varphi_{\gamma'}^*(\{\textbf{q}^x\})\psi_{\gamma}(\{\textbf{g}^x\})\sum_{\pi \in S_V}	C_\pi(\vec{\textbf{q}},\vec{\textbf{g}}),
\end{split}
\end{equation}
with
\begin{equation}\begin{split}
C_\pi(\vec{\textbf{q}},\vec{\textbf{g}}) \coloneqq  \bra{0}\prod_x[\phi(\textbf{q}^x), \phi^\dagger(\textbf{g}^{\pi(x)})]\ket{0},
\end{split}
\end{equation} 
where the vector notation $\vec{\textbf{g}}$ refers to a set of vertex variables: $\vec{\textbf{g}}=\textbf{g}^1...\textbf{g}^V$.
This formula makes explicit how the commutation properties of the field operators ensure that all the contributions coming from the various possible vertex-labellings are taken into account in the computation of the scalar product. At a combinatorial level, this means that  every vertex $x$ of one (labelled) graph overlaps with any vertex $\pi(x)$ of the other. In other words, each term $C_\pi(\vec{\textbf{q}},\vec{\textbf{g}})$ corresponds to a possible overlap configuration between two labelled versions of $\Gamma$ and $\Gamma'$.\\
We note that $C_\pi(\vec{\textbf{q}},\vec{\textbf{g}})$ corresponds to a particular ordering of the ladder operators, and exploit this observation to write the combinatorial scalar product of Definition \ref{comb1} in a second-quantized form. We start by defining the following ordering prescription:
\begin{equation}\label{pre}
:\prod_x\phi(\textbf{q}^x)\prod_x \phi^\dagger(\textbf{g}^x):_\pi \coloneqq  \prod_x\phi(\textbf{q}^x) \phi^\dagger(\textbf{g}^{\pi(x)})
\end{equation}
We then point out that the unlabelled-graph state of Eq.~\eqref{statefock} can be seen as the result of acting on the vacuum state with the operator
\begin{equation}
O_{\psi_{\Gamma}}=\int \prod_{x} \diff \textbf{g}^x \psi_\gamma(\textbf{g}^1,...,\textbf{g}^V) \prod_x  \phi^\dagger(\textbf{g}^x)
\end{equation}
In other terms, the information about the unlabelled-graph state can be equivalently encoded in an operator. By this line of reasoning, the Fock space scalar product can be seen as the vacuum expectation value of an observable constructed out of the states to be compared:
\begin{equation}\begin{split}
\langle\varphi_{\Gamma'}|\psi_\Gamma\rangle=\bra{0}O^\dagger_{\varphi_{\Gamma'}}O_{\psi_\Gamma} \ket{0} \quad .
\end{split}
\end{equation}
We might thus be tempted to define the combinatorial scalar product between two unlabelled-graph states as the vacuum expectation value of an ordered version of that observable, using the prescription introduced in Eq.~\eqref{pre}:
	\begin{equation}\begin{split}\label{comb2}
	\langle \varphi_{\Gamma'}|\psi_{\Gamma}\rangle_{\text{comb}} \overset{?}{\coloneqq}& \bra{0}:O^\dagger_{\varphi_{\Gamma'}}O_{\psi_\Gamma}:_{\bar{\pi}} \ket{0}
	\end{split}
	\end{equation}
	where the permutation $\bar{\pi}$ is such that 
\begin{equation}
|\bra{0}:O^\dagger_{\varphi_{\Gamma'}}O_{\psi_\Gamma}:_{\bar{\pi}} \ket{0}|=\max_{\pi\in S_\text{min}}|\bra{0}:O^\dagger_{\varphi_{\Gamma'}}O_{\psi_\Gamma}:_{\pi} \ket{0}| 	
\end{equation}
with $S_\text{min}$ defined in Eq.~\eqref{Sd}. A first drawback of Eq.~\eqref{comb2} is that it crucially depends on the form in which the unlabelled graph states (and so the corresponding observables) are expressed. But, more importantly, it selects a \virg{preferred} vertex-labelling and thus leads out of the Fock space; therefore, it  \textit{cannot be} the scalar product induced by the Fock space on a given subset of states. These considerations makes it clear that an alignment prescription between graphs in the Fock space is prevented by the very definition of this space, i.e. by the vertex-label symmetry underlying it. Let us stress that we do not see this as a shortcoming, but as a feature of the formalism, which correctly indicates that the only physical information is to be label-independent, and that there is no special physical reason, in this context, to partition the Hilbert space into sectors associated to different combinatorial patterns. The situation changes if new physical ingredients are introduced, leading to a meaningful, i.e. physically characterized, labelling of the vertex set. We outline a situation in which this is the case, in the following section.

\section{Effective distinguishability of vertices}\label{dist}
In a fundamental quantum gravity theory that possesses the same symmetries of classical General Relativity (even when not arising from its straightforward quantization), the only allowed reference frames are \virg{physical rods and clocks}; in other words, the presence of a background structure respect to which define a notion of space/time locality is \textit{a priori} excluded. This has led to the formulation of the relational strategy for the construction of diffeomorphism invariant observables in quantum gravity \cite{Rovelli_2002,Dittrich_2006,Dittrich_2007,Tambornino_2012, Bojowald_2011, PhysRevD.83.125023,Hoehn:2020epv}\footnote{The issue of defining and formulating physics in terms of quantum reference frames defined by suitable matter systems is also an important topic in the foundations of quantum mechanics, beside quantum gravity applications \cite{Castro-Ruiz:2019nnl,Giacomini:2017zju,Giacomini:2019fvi}.}. In the same spirit, here we show how to attain an effective distinguishability of vertices by introducing in the theory additional degrees of freedom, interpreted as discretized (scalar) matter, and to be used as a \virg{physical reference frame}, without breaking the fundamental symmetries of the formalism; in particular, the symmetry under permutations of vertex labels that we suggested as a discrete analogue of diffeomorphism invariance. Operationally, we use these additional degrees of freedom to break the symmetry over the vertex-labels at an effective level only, achieving distinguishability only for a special class of quantum states and in a physically motivated approximation.  For simplicity we consider, as additional degrees of freedom, a minimally coupled free massless scalar field $\lambda$ discretized along the geometric data on the graphs (and simplicial complexes) associated to GFT quantum states, in analogy with the approach of \cite{Oriti_2016,marchetti2020effective} for defining a relational dynamics in the GFT condensate cosmology, and based on the analysis of scalar matter coupled to quantum gravity in GFT \cite{Li:2017uao} and canonical LQG and spin foam models \cite{Domagala:2010bm, Kisielowski:2018oiv}. 
\\ The GFT field thus turns into $\phi(\textbf{g},\lambda) \in L^2(G^d/G \times \mathbb{R})$, and the canonical commutation relations of Eq.~\eqref{fockcomm} are modified as follows:
\begin{equation}\label{comm}
[\phi(\textbf{g}^x;\lambda_x), \phi^\dagger(\textbf{g}^y;\lambda_y)]=\int \diff h\prod_{i=1}^d \delta (h \text{g}^x_i\text{g}_i^{y-1}) \delta(\lambda_x-\lambda_y)
\end{equation}
For simplicity, we consider the simpler case of graph states with individually weighted vertices (the generalization to a non-separable graph wavefunction is straightforward). With the new dynamical variables given by the values of the scalar field $\lambda$, the unlabelled-graph state takes the following form:
\begin{equation}\begin{split}
\ket{\psi^{\vec{f}}_{\Gamma}}=\int \prod_{x} \diff \lambda_x \diff \textbf{g}^x\int \prod_{x<t_i(x)}\diff \textbf{h}^{x\textbf{t}(x)} \prod_{x}f_x(\textbf{g}^{x}\textbf{h}^{\pi(x) \textbf{t}(\pi(x))};\lambda_x)\prod_{x}	\phi^\dagger(\textbf{g}^x;\lambda_x)\ket{0}
\end{split}
\end{equation}
The scalar product between two graph states of this type is thus given by
\begin{equation}
\langle \psi^{\vec{f'}}_{\Gamma'}| \psi^{\vec{f}}_{\Gamma}\rangle= \sum_{\pi \in S_V}\int   \prod_x f'_{x}(\textbf{g}^{x} \textbf{h'}^{x \textbf{t'}(x)};\lambda'_{x}) f_{\pi(x)}(\textbf{g}^{x}\textbf{h}^{\pi(x) \textbf{t}(\pi(x))};\lambda'_{x}) ,
\end{equation}
where we used the commutation relations of Eq.~\eqref{comm}. \\ We then make the following assumption: in a partial semiclassical limit of the theory, 
the vertex wavefunctions are peaked on non-equal values of the scalar field $\lambda$ taken from the set $\{\lambda^0_1, ..., \lambda^0_V\}$. Then the scalar field labels can be interpreted as defining an effective embedding of the abstract graphs to which the quantum states are associated (more precisely, of their vertices) into an auxiliary manifold; but more generally, they provide a physical (i.e. in terms of measurable quantities) way to distinguish the vertices in the associated graphs. As an example, consider the case in which $f_x,f'_x$ are picked on $\lambda^0_x$; the main contribution to the scalar product then comes from the trivial permutation $\pi(x)=x$:
\begin{equation}
\langle \psi^{\vec{f'}}_{\Gamma'}| \psi^{\vec{f}}_{\Gamma}\rangle\approx \int   \prod_x f'_{x}(\textbf{g}^{x} \textbf{h'}^{x \textbf{t'}(x)};\lambda^0_x) f_{x}(\textbf{g}^{x}\textbf{h}^{x \textbf{t}(x)};\lambda^0_x)
\end{equation} 
More generally, if $f'_x$ is peaked on $\lambda^0_{\omega'(x)}$ and $f_x$ is peaked on $\lambda^0_{\omega(x)}$, where $\omega,\omega' \in S_V$, the permutation $\pi$ providing the main contribution to the scalar product is the one which satisfies the condition $\omega'(x)=\omega(\pi(x))$.
We therefore see that, if the vertex wavefunctions are peaked on values of the field taken from a discrete set $\{\lambda^0_1, ..., \lambda^0_V\}$, the scalar product is effectively performed on two labelled versions of the original unlabelled graph states, and the set $\{\lambda^0_1, ..., \lambda^0_V\}$ represents the effective vertex-labelling. As we noted already, peaking the wave-functions on $\{\lambda^0_1, ..., \lambda^0_V\}$ can be interpreted as \textit{embedding} the graph in an auxiliary manifold but in terms of physically measurable quantities, thus justifying the resulting distinguishability.

It is important to stress that the recovered distinguishability is \textit{effective}, obtained through a suitable choice of states and only, therefore, in a suitable approximation of the fundamental theory, and \textit{relational}, since it allows to align graph structures with respect to each other, as desired. In fact, we remark again that we cannot restore distinguishability of vertices at a structural level, as this is prevented from the very symmetry structure of the Fock space, and this impossibility is well grounded in the requirement of background independence of the fundamental theory.\medskip

\section{The quantum information tool of Tensor Networks}\label{TN}

A tensor $T_{n_1...n_N}$ is an array of complex numbers: the indices $n_i$ take values in a discrete set, whose dimension $D_i$ is usually called \textit{bond dimension}; the number $N$ of indices is called \textit{rank} of the tensor. Each index $n_i$ can be thought of as labelling a basis in a Hilbert space $\h_{D_i}$, and the tensor can then be regarded as a map between the Hilbert spaces associated to complementary set of indices. As an example, consider the rank-two tensor  $T_{ab}$ with input index $a$ and output index $b$; denoting by $\h_\text{in}$ and $\h_\text{out}$ the corresponding input and output Hilbert spaces, we can interpret the tensor as the following map~\cite{Pastawski_2015}: 	
	\begin{equation}
	\begin{split}
	T: \quad \h_\text{in}~~ &\rightarrow \quad \h_\text{out}\\
	\ket{a}\quad &\rightarrow \quad \sum_{b} T_{ab}\ket{b}
	\end{split}
	\end{equation}
	When regarding all the indices of a tensor $T_{n_1...n_N}$ as output indices, that tensor accounts for the state of a quantum system described by the Hilbert space  $\h_{D_1}\otimes ... \otimes \h_{D_N}$:
\begin{equation}\ket{T}= \sum_{n_1...n_N} T_{n_1...n_N}\ket{n_1...n_N}\end{equation}
 A \textit{tensor network} is a set of tensors connected according to a certain pattern, where the connection is realized by the contraction of their indices. By representing a tensor as a node with open links, one for each index, the tensor network acquires the structure of a graph. In the following, we present two of the most common tensor network structures: matrix product states (MPS) and projected entangled-pair states (PEPS).

\subsection{Matrix product states}

The matrix product state for a system of spins $s$ can be constructed by applying the Wilson renormalization group method \cite{RevModPhys.47.773,Weichselbaum_2009}, as we are going to explain (see \cite{Cirac_2009} for a detailed presentation of the topic). Let $\h_1$ be the single spin Hilbert space, having dimension $d_1=2s+1$. Given two spins $s_1$ and $s_2$, consider a subspace $\h_2 \subset \h_1 \otimes \h_1$ with $d_2\le d_1^2$. Proceed iteratively by adding spins and taking the subspace $\h_i\subset \h_{i-1}\otimes \h_1$ such that $d_i \le d_{i-1}d_1$. A state of $N$ spins in the subspace $\h_N$ can then be written as follows:
\begin{equation}\label{mps}
\ket{\psi}=\sum_{s_1,...,s_N} A^{s_1}_1A^{s_2}_2...A^{s_N}_N \ket{s_1,s_2,...,s_N}
\end{equation}
where
$A^s_i\in M_{d_{i-1} \times d_i}$ for $i=2, ..., N-1$, $A^s_1$ is a row vector of rank $d_1$ and $A^s_N$ is a column vector of rank $d_N$. Explicitly:
\begin{equation}
\ket{\psi}=\sum_{s_1,...,s_N} (A^{s_1}_1)_{\alpha}(A^{s_2}_2)_{\alpha\beta}...(A^{s_{N-1}}_{N-1})_{\mu\nu}(A^{s_N}_N)_{\nu} \ket{s_1,s_2,...,s_N},
\end{equation}
where $\alpha, \beta, ..., \nu$ are the matrix indices. To each site $i$ we thus associated a tensor $(A^s_i)_{\alpha\beta}$ that, in addition to the physical index $s_i$, has left and right virtual indices  $\alpha$ and $\beta$ connecting it with sites $i-1$ and $i+1$, respectively. The virtual indices can be thought of as describing the states of auxiliary systems added to each site.

\subsection{Projected entangled-pair states}

The MPS of Eq.~\eqref{mps} can be expressed as another type of tensor network decomposition, called projected entangled-pair states (PEPS). 
Let $|\epsilon_i\rangle\in \h_{i}\otimes \h_{i}$ be a maximally entangled state between the right ancilla of site $i$ and the left ancilla of site $i+1$. Consider the operator $P_i : \h_{i-1}\otimes \h_{i}\rightarrow \h_1$ projecting the ancilla states to the $s$-spin state:
\begin{equation}
P_i=\sum_s \sum_{\alpha\beta} (A^s_i)_{\alpha\beta}\ket{s}\langle\alpha\beta| 
\end{equation}
The matrix product state of Eq.~\eqref{mps} can then be written as follows:
\begin{equation}\label{PP}
\ket{\psi}=\sum_{s_1,...,s_N} A^{s_1}_1A^{s_2}_2...A^{s_N}_N \ket{s_1,s_2,...,s_N}=P_1 \otimes ... P_N |\epsilon_1\rangle\otimes...|\epsilon_{N-1}\rangle
\end{equation}
Such a tensor network decomposition is thus built up from maximally entangle ancilla pairs (making up the links of the network) which are projected to physical spins (the network sites). Here we considered a one-dimensional system, a spin chain, but the PEPS decomposition can be easily generalised to higher dimensional systems. Before illustrating this point, we show that in a simple one-dimensional PEPS, specifically the state of a transitionally invariant system, entanglement entropy is bounded by an area law. To start note that, since the reduced density matrix $\rho_{1,2,...,M}$ for the first $M$ spins has rank bounded by $d_M$, the entanglement entropy $S(\rho_{1,2,...,M})$ satisfies
\begin{equation}
S(\rho_{1,2,...,M})\le \log d_M.
\end{equation}
We assume that all the auxiliary systems have dimension $D$, and site $N$ is connected to site $1$, i.e. $A^s_i\in M_{D \times D}$ $\forall i$. The state of the translationally invariant spin chain thus takes the following form:
\begin{equation}
\ket{\psi}=\sum_{s_1,...,s_N}\text{Tr}\left(A^{s_1}_1A^{s_2}_2...A^{s_N}_N\right) \ket{s_1,s_2,...,s_N}.
\end{equation}
Given an interval $A$ of the spin chain, the entanglement entropy $S_A$ therefore satisfies
\begin{equation}
S_A\le 2 \log D ,
\end{equation}
which is an area law. In fact, denoting by $c_A$ the curve bounding $A$, we have that
\begin{equation}
S_A\le \min_{c_A}\{ c_A\cap \text{network} \} \log D,
\end{equation}
where $\{ c_A\cap \text{network} \}$ is the number of intersections between the curve $c_A$ and the spin chain.

We can introduce PEPS in dimension higher than one by considering network sites that, in addition to the physical spin $s$, have an arbitrary number of auxiliary spins which are maximally entangled with their neighbours. For simplicity we assume that each auxiliary spin has dimension $D$. The PEPS is then constructed by projecting the entangled auxiliary-spin pairs onto the physical spins with the following operators:
\begin{equation}\label{PEPS}
P_i=\sum_s \sum_{\alpha\beta\gamma...=1}^D (A^s_i)_{\alpha\beta\gamma...}\ket{s}\langle\alpha\beta\gamma...| ,
\end{equation}
where $\alpha\beta\gamma...$ are the virtual indices referring to the auxiliary spins.
Consider now a PEPS in which all auxiliary spins have the same dimension $D$; let $A$ be a region of the network and $c_A$ its boundary.  Since every maximally entangled state between an auxiliary spin and its neighbour has dimension $D$, we have that
\begin{equation}
S_A\le \min_{c_A}\{ c_A\cap \text{network} \} \log D,
\end{equation}
where $\{c_A\cap \text{network}\}$ corresponds to the number of entangled pairs across the boundary $c_A$ of region $A$. Therefore, also in this case the entanglement entropy turns out to be bounded by an area law. Note that an area law bound for the entanglement entropy arises in tensor network directly from their definition, as the contraction of indices generally induces entanglement between the corresponding degrees of freedom. However, only some tensor networks saturate this bound, thus exhibiting an holographic behaviour; among them, we can find tensor networks built up from perfect tensors (a special class of isometric tensors) \cite{Pastawski_2015} or random tensor network in the limit of large bond dimensions \cite{Hayden_2016}.

It is possible to construct PEPS with completely arbitrary network geometries by varying the number of auxiliary spins of each site, and defining their entanglement relations by choosing appropriate ancilla-pair states. See \cite{Qi_2017} for an example of such construction: the vertices possess the highest possible valence for a completely connected graph (namely $V-1$ for a graph of size $V$), and separable ancilla states are introduced to account for the absence of links between vertices.

PEPS are used for the study of lattice gauge theories (LGT) through tensor networks techniques \cite{Tagliacozzo_2011,Tagliacozzo_2014}. In particular, in the LGT context they are provided with a gauge-invariance symmetry at each node, thus resembling also in this aspect the structure of GFT graphs. The second-quantized tensor networks that we are going to define in the GFT context can indeed be seen as a generalization of such construction.

\section{A dictionary between Group Field Theory states and (generalised) Tensor Networks}\label{dict}

\subsection{GFT and TN: a simple realization of entanglement/topology and entanglement/geometry correspondences}

Before presenting the map between group field theory states and tensor networks, we highlight some features which are highly relevant from a quantum gravity perspective. 

\begin{itemize}
\item[-] \textit{Entanglement/connectivity correspondence} - A first one is the relation between entanglement and connectivity of the network/graph. As previously explained, both frameworks employ entanglement as the glue of these structures. In the GFT context, due to the simplicial interpretation of the graph, this feature implies a relation between entanglement and connectivity of space; in fact, links made of entangled vertex-lines correspond to adjacency relations of the cells dual to the involved vertices. To make clearer the role played by entanglement in the connectivity of a GFT graph structure, let us focus on a very simple example: two vertices connected by a link, where the latter is made of vertex-lines of different colours, say $a$ and $b$. In spin representation, the gluing of two vertex-lines corresponds to the contraction of the labels at their endpoints, here indicated by $n^1_a$ and $n^2_b$:
	\begin{equation*}\begin{split}
	\int \diff h\phi^\dagger(\textbf{g}^1;g^1_a h)\phi^\dagger(\textbf{g}^2;g^2_b h) \ket{0}=&\sum%_{\{\text{j}^1_{i\neq a}\}\{\text{j}^2_{i\neq b}\}}
	\left(\sum_{n} \phi^{\dagger\textbf{j}^1 \text{j}\iota^1}_{\textbf{n}^1;n^1_a=n} \phi^{\dagger\textbf{j}^2\text{j}\iota^2}_{\textbf{n}^2;n^2_b=n}\right) C^{\textbf{j}^1\text{j}\iota^1}_{\textbf{m}^1;m^1_a}C^{\textbf{j}^2\text{j}\iota^2}_{\textbf{m}^2;m^2_b}\\&\hspace{2cm}\cdot \prod_{i\neq a}d_{j^1_i} D^{j^1_i}_{m^1_i n^1_i}(g^1_i )\prod_{i\neq b} d_{j^2_i} D^{j^2_i}_{m^2_i n^2_i}(g^2_i )~d_{j^1_a}D^{\text{j}}_{m^1_a m^2_b}(g^1_a g^{2-1}_b)\ket{0}
	\end{split}
	\end{equation*}
	where we made explicit the sum regarding the open ends to be glued (also called \virg{semi-links} in the following), while we grouped in the first summation-sign that over all the other repeated indices. The expression inside the round brackets clearly shows that the gluing process corresponds to the formation of an entangled state of the degrees of freedom associated to the semi-links involved. Every link in the GFT graph is therefore an \virg{entanglement link}; in particular, a link $\ell$ carrying the spin $j$ corresponds to the maximally entangled state of the semi-links forming it:
	\begin{equation}\label{mes}
\ket{\ell}=\frac{1}{\sqrt{d_{j}}}\sum_n  \ket{j; n }\otimes \ket{j;n}
\end{equation}
This means that, since entanglement controls the connectivity of a graph, it determines the topology of the simplicial complex dual to it. We have, therefore, an explicit example of an entanglement/topology correspondence.\\
In the tensor network context, a simplicial-geometry interpretation of the network is possible when the latter is proved to reproduce a discretized manifold, as it happens for tensor networks modelling AdS/CFT states \cite{Qi_2017, Hayden_2016}. There is however a crucial difference respect to the GFT case: in the mentioned tensor network constructions, the geometric interpretation is induced \virg{at a later stage}, by defining a metric through the graph distance. We showed that for GFT graphs, instead, the geometric characterization arises naturally thanks to the presence, on top of the combinatorial structure, of additional quantum geometric degrees of freedom.\\
In quantum gravity, a link between entanglement and space(time) connectivity has been clearly pointed out, for example, in the cited work \cite{articleRaa}, where it was shown, by a thought experiment in the AdS/CFT context, that disentangling two sets of degrees of freedom in the CFT corresponds to increasing the proper distance between the dual spacetime regions, while the area separating them decreases.

\noindent This is the combinatorial and topological side of the story. In fact, there is an additional geometric side of the same story, which is particularly interesting from the point of view of quantum gravity (including the GFT formalism and beyond it, in AdS/CFT applications, LQG etc.): the entanglement so established carries a straighforward geometric interpretation, and corresponding entanglement measures can be seen to be measuring geometric observables.

\item[-] \textit{Primitive entanglement/area correspondence} - In the geometric interpretation of spin network graphs in the context of GFT (and LQG), a link of the graph is dual to a surface, i.e. a portion of surface on the shared boundary of the two polyhedra (simplices, in the case we considered) dual to the two vertices sharing the link, and the spin attached to it labels the eigenvectors of the area operator associated to that surface. The spectrum of the area operator for such dual surface is (using symmetric ordering) $\sqrt{j(j+1)}$ in Planck units, and thus it scales like $j$ for largish eigenvalues. This is also the scaling of the dimension of the Hilbert space of states associated to each link labeled by a given spin, i.e. a maximally entangled state, which is $dim(j) = 2 j + 1$. In turn, this dimension gives a simple measure of the entanglement that we have seen being associated to the same link, thus establishing a sort of \virg{primitive entanglement/area correspondence} in our quantum gravity states.  

\item[-] \textit{Primitive entanglement/volume correspondence} - An entanglement process can be identified as lying also at the origin of the intertwiner degrees of freedom, which are attached to the vertices of the graph associated to GFT states (and thus to the tensors of the corresponding tensor networks). In fact, the intertwiner arises from the \virg{gluing} of open lines into a vertex, by means of the requirement of local gauge invariance. The spin network wave-function (defined in Eq.~\eqref{spindef}) can indeed be decomposed as follows: 
\begin{equation}\begin{split}\label{iota}
\psi_{\textbf{j}\textbf{n}\iota}(\textbf{g})=&\sum_{p_1...p_d}C^{j_1...j_d \iota}_{p_1...p_d} \prod_i \sqrt{d_{j_i}} D^{j_i}_{p_i n_i}(g_i)\\=&\bra{\textbf{g}}\sum_{p_1...p_d}\left(\bigotimes_i\sqrt{d_{j_i}} \ket{j_i; n_i}\bra{j_i; p_i }\right)\left(\sum_{p'_1...p'_d}C^{j_1...j_d \iota}_{p'_1...p'_d}\ket{j_i;p'_1} \otimes ... \otimes \ket{j_d; p'_d}\right)
\end{split}
\end{equation}
The second line of Eq.~\eqref{iota} shows that $\psi_{\textbf{j}\textbf{n}\iota}(\textbf{g})$ can be seen as the result of contracting line states (round brackets on the left) with an entangled state of (equal-side) open ends of that lines (round brackets on the right). This is one more instance of a straightforward entanglement/geometry correspondence at the discrete (simplicial) geometry level. In fact, the entanglement structure is controlled by the degree of freedom $\iota$, the intertwiner quantum number. This, in turn, can be shown (in both simplicial quantum geometry, GFT and LQG) to label eigenvalues of the operator measuring the volume of the polyhedron dual to the spin network vertex. Thus, also volume information is a measure of the entanglement of quantum gravity degrees of freedom. 

\item[-] \textit{Entanglement/area laws} - A well known consequence of the entanglement origin of tensor networks is the fact that, as showed for the translationally invariant examples presented in the previous section, the entanglement entropy is bounded by an area law: given a region $A$ of the network bounded by the curve $c_A$, and denoted by $D$ the dimension of the Hilbert space associated to the links, we have that
\begin{equation}\label{area}
S_A\le \min_{c_A}\{ c_A\cap \text{network} \} \log D
\end{equation}
When interpreting $\log D$ as the area of an elementary surface dual to the network link,  Eq.~\eqref{area} turns into an area law for the upper bound to the entanglement entropy. In fact, $\{ c_A\cap~\text{network} \}$ counts the number of intersections between the boundary $c_A$ and the network, i.e. the number of surface units in $c_A$, and $\{ c_A\cap \text{network} \}\log D$ thus provides the area of the boundary surface $c_A$. Entanglement area laws are of great interest in quantum gravity, since the latter is expected to exhibit an holographic behaviour, as suggested by the scaling of black hole entropy with the horizon area and the Ryu-Takayanagi formula~\cite{PhysRevLett.96.181602,Ryu_2006}, which relates the entanglement entropy in $\text{CFT}_{d+1}$ to the area of $d$-dimensional minimal surfaces in the dual $\text{AdS}_{d+2}$. For tensor networks modelling holographic states in the AdS/CFT correspondence (as in \cite{swingle2012constructing}, where the tensor network arises by entanglement renormalization, and in \cite{Bao_2019}, where it is constructed by entanglement distillation) Eq.~\eqref{area} acquires precisely the connotation of an area law for the entanglement entropy. 
Since in GFT states the spins carried by a link are eigenvalues of an area operator associated to the surface dual to it, as we mentioned, the bound to the entanglement entropy of a link, and hence of an extended region of a GFT graph, naturally have an area law interpretation\footnote{We are considering the simplest case of a graph with fixed spins, and ignoring for simplicity the contribution to the entropy deriving from the intertwiner degrees of freedom.}. GFT states therefore share with general tensor networks the feature of having an entanglement entropy bounded according to Eq.~\eqref{area}; just as there are classes of tensor networks that saturate the bound (and thus have an holographic nature), certain GFT states have proved to satisfy an entanglement area law: in~\cite{Chirco_2018}, for example, the Ryu-Takayanagi formula was derived for a GFT graph in first-quantization. Let us finally point out that the area bounding a region of the GFT complex depends on the entanglement entropy of the links crossing it, whose total number is determined by the combinatorial structure of the graph. That general area law is thus the result of the graph connectivity and of the local contributions to the entanglement entropy, in turn carrying a primitive entanglement/area correspondence.

\end{itemize}

\subsection{GFT graph states as PEPS}
We are going to show that GFT labelled-graph states can be understood as generalised PEPS and that, consequently, unlabelled ones realize an analogous correspondence in a second-quantization setting, leading to the definition of second-quantized tensor networks. 

As explained in Section \ref{TN}, PEPS are constructed by projecting maximally-entangled ancilla-pairs indices onto \virg{physical} indices (attached to the nodes of the network). In the GFT context, the role of ancilla-pair states is played by link states, i.e. maximally entangled states of edge spins, and node degrees of freedom translate into open-vertex ones. We clarify that with an example, and then present the more general case. Given a completely connected graph $\gamma$, for each link $\ell=(x,y;i)$ consider the following maximally entangled state in the Hilbert space $\h^{j^x_i=j}\otimes \h^{j^{y}_i=j}$:
\begin{equation}
\ket{\ell=(x,y;i)}=\frac{1}{\sqrt{d_{j}}}\sum_m  \ket{j; m }\otimes \ket{j;m},
\end{equation}
and for each vertex a generic state
\begin{equation}\label{T}
\ket{v_x} =\sum_{\textbf{j}\textbf{n}\iota} T^{\textbf{j} \iota}_{x;\textbf{n}} \ket{\textbf{j};\textbf{n};\iota}
\end{equation}
where $\ket{\textbf{j};\textbf{n};\iota}$ is the spin network basis: $\langle \textbf{g}|\textbf{j};\textbf{n};\iota \rangle = \psi_{\textbf{j}\textbf{n}\iota}(\textbf{g})$, with $\psi_{\textbf{j}\textbf{n}\iota}(\textbf{g})$ the spin network wavefunction defined in Eq. \eqref{spindef}. 
Then perform the contraction
\begin{equation}\label{GFTPEPS}
 \bigotimes_{\ell \in A} \bra{\ell} \bigotimes_x \ket{v_x} =\sum_{\iota^1,...,\iota^N}\Tr_A
\left(T_{1}^{ \textbf{j}^1\iota^1}... ~ T_{N}^{\textbf{j}^N \iota^N}\right) |\iota^1,...,\iota^N\rangle, %\equiv \ket{\varphi_\gamma}
\end{equation}
where $j^x_i=j~\forall x,i$, $A$ is the adjacency matrix which encodes the combinatorial pattern of $\gamma$ and $\Tr_A$ is the tensorial trace contracting the vertex tensors according to it:
\begin{equation}
\Tr_A
\left(T_{1}^{ \textbf{j}^1\iota^1}... ~ T_{N}^{\textbf{j}^N \iota^N}\right)=T_{1;\textbf{n}^1}^{\textbf{j}^1\iota^1}... ~ T_{N;\textbf{n}^N}^{\textbf{j}^N \iota^N}\prod_{a^i_{xy}=1}\delta_{n^x_i,n^y_i}
\end{equation}
The state defined by Eq.~\eqref{GFTPEPS}, associated by construction to the graph $\gamma$, is a tensor network of the form of Eq.~\eqref{PP}, where the intertwiners $\iota^x$ play the role of  \virg{physical} indices, and $n^x_{i}$ that of \virg{virtual} indices, with fixed bond dimension $d_j$; in fact, in this simple example all links carry the same spin $j$. A more general setting can be considered by taking, as GFT counterparts of the TN ancilla-pair states, link states in the direct sum of the Hilbert spaces associated to all group representations:
\begin{equation}
\ket{\ell=(x,y;i)}=\bigoplus_j\frac{1}{\sqrt{d_{j}}}\sum_m  \ket{j; m }\otimes \ket{j;m}.
\end{equation}
The tensor network resulting from the contraction defined in Eq. \eqref{GFTPEPS} is then the following:
\begin{equation}\label{1stTN}
\Tr_A
\left(T_{1}^{ \textbf{j}^1\iota^1}... ~ T_{N}^{\textbf{j}^N \iota^N}\right)=T_{1;\textbf{n}^1}^{\textbf{j}^1\iota^1}... ~ T_{N;\textbf{n}^N}^{\textbf{j}^N \iota^N}\prod_{a^i_{xy}=1}\delta_{j^x_i,j^{y}_i}\delta_{n^x_i,n^y_i}
\end{equation}
Let us now move the second-quantization framework. In particular, consider a GFT unlabelled-graph state constructed out of individually-weighted vertices, where the latter are given by Eq.~\eqref{T}:
\begin{equation}\label{2ndTNiwv}
\ket{\psi_\Gamma^{\vec{T}}}=\sum_{\vec{\textbf{j}},\vec{\textbf{n}},\vec{\iota}} \left(\sum_{A'\in [A]}\prod_x T^{\textbf{j}^{\pi(x)}\iota^{\pi(x)}}_{x;\textbf{n}^{\pi(x)}}\prod_{a'^i_{xy}=1}\delta_{j^x_i,j^y_i}\delta_{n^x_i,n^y_i}\right)\prod_{x=1}^V\phi^{\textbf{j}^x\iota^x\dagger}_{\textbf{n}^x}\ket{0}
\end{equation}
We recognize within the round brackets a tensor network which is the symmetrized version of that in Eq.~\eqref{1stTN}, and can be understood as a \textit{second-quantized tensor network}. The argument can be extended to arbitrary GFT unlabelled-graph states, which take the form
\begin{equation}
\ket{\psi_\Gamma}=\sum_{\vec{\textbf{j}},\vec{\textbf{n}},\vec{\iota}} \left(\sum_{A'\in [A]}\psi^{\textbf{j}^{\pi(1)}...\textbf{j}^{\pi(N)}\iota^{\pi(1)}...\iota^{\pi(N)}}_{\textbf{n}^{\pi(1)}...\textbf{n}^{\pi(N)}}\prod_{a'^i_{xy}=1}\delta_{j^x_i,j^y_i}\delta_{n^x_i,n^y_i}\right)\prod_{x=1}^V\phi^{\textbf{j}^x\iota^x\dagger}_{\textbf{n}^x}\ket{0}
\end{equation}
Note that this expression reduces to Eq.~\eqref{2ndTNiwv} for
\begin{equation}\label{2ndTNgen}
\psi^{\textbf{j}^{\pi(1)}...\textbf{j}^{\pi(N)}\iota^{\pi(1)}...\iota^{\pi(N)}}_{\textbf{n}^{\pi(1)}...\textbf{n}^{\pi(N)}}=\prod_x T^{\textbf{j}^{\pi(x)}\iota^{\pi(x)}}_{x;\textbf{n}^{\pi(x)}}
\end{equation}

Let us finally remark the features of GFT graph states which characterize them as \textit{generalised} tensor networks. Some of them are already present at the first-quantized level. The bond dimensions of tensor indices, i.e. the spins associated to the links, are not fixed parameters, but truly dynamical variables; in fact, strictly speaking each Hilbert space associated to a link (before additional conditions are taken into account) is infinite dimensional, being isomorphic to $L^2(G)$. Moreover, the \virg{physical} indices are not, in general, independent from the \virg{virtual} ones. Note also that, as pointed out in \cite{Chirco_2018}, already the first-quantized GFT graph states can be seen as \textit{random} tensor networks, where the randomness is defined over a probability distribution set by the GFT dynamics; this remains true at the second-quantized level. A feature which instead pertains more naturally to the second-quantization framework is the dynamical nature of the combinatorial structure: since the network arises from the dynamics of a field, vertices can be created or destroyed, and graph connectivity (deriving from the entanglement properties of the field excitations) can vary.
We also point out that, as we noted in quantum gravity applications with a simplicial-geometry interpretation, the GFT quanta are endowed with a local gauge symmetry (invariance under the diagonal action of a Lie group), which makes their quantum states corresponding to {\it symmetric} tensor networks, of the type employed in applications to gauge theories. 

\section{Discussion}
The GFT formalism describes entanglement graphs representing simplicial complexes which are understood as spatial portions of a quantum spacetime (or, more generally, codimension one submanifolds). These structures naturally satisfy a discrete version of diffeomorphism invariance, as they are symmetric respect to permutations of the vertex-labelling used to define them\footnote{Note that the links of the graph, as adjacency relations among vertices, are defined by the vertex labels themselves.}. In fact, a given vertex-labelling for an entanglement graph can be understood as a choice of coordinate system on the (discretized) spatial manifold it describes. Invariance under vertex-relabelling can thus be regarded as the discrete analogue of diffeomorphism invariance. 

Entanglement graphs have been first defined in the pre-Fock space, where distinguishability of vertices enables to define a combinatorial pattern among them, then constrained with the aforementioned symmetry. The pre-Fock and Fock spaces of the theory allow (in fact, make mandatory) to consider also superpositions of labelled and unlabelled entanglement graphs, respectively. The two are conceptually quite different. 

In the pre-Fock space of distinguishable vertices, graphs in quantum superposition can be aligned according to the given vertex labelling. In a discrete-gravity perspective, we could say that superposing labelled entanglement graphs amounts to superposing discrete metrics (to the extent in which they are encoded in the combinatorial pattern only). A notion of graph superposition has recently been provided in \cite{arrighi2020quantum} through the definition of an Hilbert space for coloured graphs, where colours are generic field data. When the latter have a geometric interpretation, that coloured graphs coincide, at a formal/descriptive level, with our labelled entanglement graphs.  At a structural level, the difference is in taking graphs as basic structures, decorated with some data \virg{at a later stage} (case of \cite{arrighi2020quantum}), or having them emergent from the quantum behaviour of a many-body system (GFT case). The first setting naturally implies an orthogonality relation among different graphs, which, instead, is not necessarily satisfied in the second: the scalar product between labelled-graph states in the GFT pre-Fock space can be non vanishing even for non equal graphs, precisely because the latter are just \textit{features} of the many-body states and, specifically, manifestations of their entanglement content. Note that, though the Hilbert spaces describing graphs in the two contexts have a different structure, a robust notion of graph superposition naturally derives from both of them. 

Once it has been established that vertex labelling does not possess any physical meaning, comparing graphs independently on it becomes particularly relevant. In \cite{arrighi2020quantum} Arrighi \textit{et al.} stress that, if vertex labels were \textit{a priori} not observable, the scalar product between coloured graphs differing only for that labels would be 1; as it is not the case (the result is actually zero) invariance under vertex relabelling must be enforced. In the GFT pre-Fock space the scalar product between isomorphic entanglement graphs, though \textit{a priori} not zero, is not necessarily equal to 1. We defined an alternative scalar product which gets such an outcome, as compares entanglement graphs with the goal of maximising their overlap, regardless of the vertex labelling.

In addition to the pre-Fock space of labelled-graph states and their superpositions, our framework includes the space of properly physical, i.e. \virg{diffeomorphism invariant}, states: the Fock space. Within it, we have naturally superpositions of unlabelled entanglement graphs, which can be understood, at a discrete-gravity level, as superpositions of geometries (i.e. equivalence classes of metrics). Note that a simple alignment prescription is not possible among unlabelled graphs, exactly as a notion of locality is not available when working with geometries. 
It could be possible, in principle, to define topological observables that capture the purely combinatorial, label-independent pattern encoded in a graph, i.e. associated to its entire equivalence class under graph isomorphisms. However, we leave this possibility for further work. 
Beside this possibility, we highlighted that a straightforward alignment prescription can be recovered when new degrees of freedom, interpreted as discretized matter, are added to the fundamental model, in the same spirit of the construction of relational (and diffeomorphism-invariant) observables in quantum gravity. In particular, we have shown that certain states allow to restore an effective (and relational) distinguishability of vertices thanks to their semi-classical behaviour with respect to the additional degrees of freedom. 

\section{Conclusions and outlook}

The complexity of the quantum gravity problem has led to a proliferation of strategies to approach it. Among them, tensorial group field theories, which are intended as a quantum field theories \textit{of} spacetime, distinguish themselves by their cross-cutting nature, given by the multiple connection with other quantum gravity approaches. In this paper, we have shown that, in group field theory, discretized spatial geometries arise as entanglement patterns among quanta of space, the excitations of the GFT bosonic field. We provided a detailed picture of the identification of such entanglement graphs among the GFT states. We exploited the distinguishability of vertices in the pre-Fock space of the theory to define in the latter a prescription for the construction of entanglement graphs with arbitrary connectivity, as well as a scalar product to compare them on the basis of their combinatorics. We then removed the unphysical vertex-labelling to implement on the entanglement graphs a discretized version of diffeomorphism invariance. An effective notion of distinguishability, needed for practical reasons, is then recovered in the semi-classical regime of an extended model comprising an additional degree of freedom playing the role of a reference frame. Finally, we showed that GFT entanglement graphs match well known quantum information structures, with a high computational efficiency: tensor networks. In doing that, we generalised to the second-quantization setting the intuition of \cite{Chirco_2018}, implemented in first-quantization. 
A different reading of this correspondence is that, once transposed in the GFT framework, tensor networks inherit a simplicial-geometry interpretation and a second-quantized model characterization.

Such a dictionary paves the way for exploiting in a much more intensive way tensor networks techniques in quantum gravity calculations. 

To give an example, since graphs correspond, in the GFT context, to patterns of entanglement, TN operations (such as disentanglers and coarse-grainers \cite{swingle2012constructing}) could be used to define observables capable to extract the combinatorial-pattern information from GFT states. 

Moreover, as a wide class of tensor networks (for example, built from random \cite{Qi_2017,Hayden_2016} or perfect \cite{Pastawski_2015} tensors) exhibits an holographic behaviour, we expect the established correspondence with GFT states to simplify the study of holographic properties of the latter. In particular, we have in mind the generalization to the GFT framework of recent works that investigate the relation between bulk and boundary degrees of freedom of random tensor networks, by regarding the latter as maps between such degrees of freedom (see for example \cite{Qi_2017,Hayden_2016}); defining a similar map for GFT states, by taking the intertwiners as bulk degrees of freedom, will make possible to study how volume correlations (entanglement among intertwiners) affects the properties of the graph boundary. We expect this bulk entanglement to provide corrections to the  Ryu–Takayanagi formula (recovered in the GFT context in \cite{Chirco_2018}), in analogy to what is the case for random tensor networks \cite{Hayden_2016}. 

In this programme, noteworthy from a quantum gravity perspective is the idea underpinning GFT graphs and distinguishing them from (random) tensor networks: the GFT structures are not just (background) structures decorated with some labels, but manifestations of the interaction of degrees of freedom with a genuine geometrical interpretation, whose \virg{randomness} is induced by the GFT model which determines their dynamics. Moreover, our dictionary will allow to translate the aforementioned results to a second-quantized (and hence diffeomorphism-invariant, in a discrete quantum gravity interpretation) language.

Finally, we would like to remark that the potential of our dictionary relies on the fact that, to extract continuum physics from the GFT formalism, we need to control the regime of the theory with a large number of interacting quanta, and tensor networks can efficiently tackle such a computational problem.  Possible candidates for states modelling an effective continuum-geometry are condensate states \cite{Oriti_2015,Oriti_2016}, and our dictionary could be used to analyse them in a quantum-information theoretic setting, and to characterize them in terms of information-transmission  properties. An important application concerns the GFT condensate states introduced for modelling quantum black holes \cite{PhysRevLett.116.211301, PhysRevD.97.066017}: the aforementioned strategy could in fact be applied to characterize the event horizon in information-theoretic terms, before looking at the translation of such characterization in geometric terms.
The usefulness of the correspondence goes in the opposite direction too. A number of results and techniques developed in the context of random tensor models and tensorial group field theories can be useful in standard quantum many-body systems, improving standard tensor networks applications. We have in mind in particular the body of work on large-N expansions and universality results \cite{Gurau_2012}.

\section*{Acknowledgments}
The authors would like to thank Goffredo Chirco, as well as the quantum gravity group at LMU, for useful discussions and comments. EC acknowledges financial support from the European Research Council (ERC) under the Starting Grant GQCOP (Grant No. 637352), and thanks Gerardo Adesso for encouragement and support. EC also thanks the Ludwig Maximilian University of Munich for the hospitality.
\appendix
\section{Scalar product between graph basis states} \label{scalarp}
Here we want to compute the scalar product between graph basis states in spin representation. We start by rewriting the graph basis wave-functions in the following form:
\begin{equation}\begin{split}
\theta_{\gamma \{n^x_i\}_{\text{open}}}^{\{j_i^{xt_i(x)}\}\vec{\iota}}(\{g_\ell=g^x_i g^{t_i(x) -1}_i\})=C^{\{j_i^{xt_i(x)}\}\vec{\iota}}_{\vec{\textbf{m}}}\left( \prod_{x,i:x<t_i(x)} \sqrt{d_{j^{xt_i(x)}_i}} \right)\prod_x  D^{j^{xt_i(x)}_i}_{m^x_i q^{xt_i(x)}_i}(g^x_i )
\end{split}
\end{equation}
where the labels $q_i^{xt_i(x)}$ are such that $q^{xy}_i=q^{yx}_i$ (the ones corresponding to internal links are thus summed over) and $q^{x0}_i=n^x_i$. We then have that
\begin{equation}\begin{split}
\langle\theta_{\gamma'}(\{j'_\ell,n'_\ell,\vec{\iota'}\})|\theta_\gamma(\{j_\ell,n_\ell,\vec{\iota}\})\rangle =& \int   \diff g^x_i\theta_{\gamma' \{n'\}_{\text{open}}}^{\{j_i'^{xt'_i(x)}\}\vec{\iota'}}(\{g^{xt'_i(x) }_i\})\theta_{\gamma^\pi \{n\}^\pi_{\text{open}}}^{\{j_i^{xt_i(x)}\}\vec{\iota}}(\{g^{xt_i(x) }_i\})\\=& C^{\{j_i'^{xt'_i(x)}\}\vec{\iota'}}_{\vec{\textbf{m'}}}~C^{\{j_i^{xt_i(x)}\}\vec{\iota}}_{\vec{\textbf{m}}}\prod_\ell \sqrt{d_{j_\ell}}\prod_{\ell'} \sqrt{d_{j_{\ell'}}}\prod_{x,i} \int \diff g^x_i D^{j'^{xt'_i(x)}_i}_{m'^x_i q'^{xt'_i(x)}_i}(g^x_i ) D^{j^{xt_i(x)}_i}_{m^x_i q^{xt_i(x)}_i}(g^x_i )\\=&C^{\{j_i'^{xt'_i(x)}\}\vec{\iota'}}_{\vec{\textbf{m'}}}~C^{\{j_i^{xt_i(x)}\}\vec{\iota}}_{\vec{\textbf{m}}}\prod_\ell \sqrt{d_{j_\ell}}\prod_{\ell'} \sqrt{d_{j_{\ell'}}}\prod_{x,i} \delta_{j_i'^{xt'_i(x)},j^{xt_i(x)}_i} \delta_{m'^x_i,m^x_i}\delta_{q'^{xt'_i(x)}_i,q^{xt_i(x)}_i}~D%\\=&\prod_\ell \sqrt{d_{j_\ell}}\prod_{\ell'} \sqrt{d_{j_{\ell'}}}\prod_{x,i} \delta_{j_i'^{xt'_i(x)},j^{xt_i(x)}_i} \delta_{q'^{xt'_i(x)}_i,q^{xt_i(x)}_i}~\delta(\vec{\iota},\vec{\iota'})~D\\=&\prod_x \delta_{j_i'^{xt'_i(x)},j^{xt_i(x)}_i}\prod_{x:t_i(x)= 0, t'_i(x)\neq 0}\delta_{n^{t'_i(x)}_i,n^{x}_i}\prod_{x:t_i(x)\neq 0, t'_i(x)= 0}\delta_{n'^{x}_i,n'^{t_i(x)}_i}\\ &\cdot\prod_{x:t_i(x)=t'_i(x)= 0}\delta_{n'^{x}_i,n^{x}_i}  ~\delta(\vec{\iota},\vec{\iota'})
\end{split}
\end{equation}
where  
\begin{equation}
D\coloneqq\prod_{x,i:t_i(x)\neq 0, t'_i(x)\neq 0}\frac{1}{(2j^{xt_i(x)}_i+1)}\prod_{x:t_i(x)= 0 \wedge t'_i(x)= 0}\frac{1}{(2j^{x}_i+1)} .
\end{equation}
By using the relation
$C^{\{j\}\vec{\iota'}}_{\vec{\textbf{m'}}}~C^{\{j\}\vec{\iota}}_{\vec{\textbf{m}}}\prod_x \delta_{m'^x_i,m^x_i}=\delta(\vec{\iota},\vec{\iota'})$ we finally obtain
\begin{equation}\begin{split}
\langle\theta_{\gamma'}(\{j'_\ell,n'_\ell,\vec{\iota'}\})|\theta_\gamma(\{j_\ell,n_\ell,\vec{\iota}\})\rangle =&\prod_\ell \sqrt{d_{j_\ell}}\prod_{\ell'} \sqrt{d_{j_{\ell'}}}\prod_{x,i} \delta_{j_i'^{xt'_i(x)},j^{xt_i(x)}_i} \delta_{q'^{xt'_i(x)}_i,q^{xt_i(x)}_i}~\delta(\vec{\iota},\vec{\iota'})~D\\=&\prod_x \delta_{j_i'^{xt'_i(x)},j^{xt_i(x)}_i}\prod_{x:t_i(x)= 0, t'_i(x)\neq 0}\delta_{n^{t'_i(x)}_i,n^{x}_i}\prod_{x:t_i(x)\neq 0, t'_i(x)= 0}\delta_{n'^{x}_i,n'^{t_i(x)}_i}\\ &\cdot\prod_{x:t_i(x)=t'_i(x)= 0}\delta_{n'^{x}_i,n^{x}_i}  ~\delta(\vec{\iota},\vec{\iota'}).
\end{split}
\end{equation}

\bibliographystyle{jhep}
\bibliography{graphbib}
\end{document}

%% file: def_new.tex
\usepackage{graphicx}
\usepackage{amsmath}
\usepackage{mathrsfs}
\usepackage{dsfont}
\usepackage{amssymb}
\usepackage{subfigure}
\usepackage{color}
\usepackage{blkarray}
\usepackage[breaklinks=true,colorlinks=true,linkcolor=blue,urlcolor=blue,citecolor=blue]{hyperref}
\def\bra#1{{\left\langle #1 \right|}}
\def\ket#1{{\left| #1 \right\rangle}}

\definecolor{amber}{rgb}{1.0, 0.49, 0.0}
\usepackage[dvipsnames]{xcolor}

\usepackage{verbatim}
\usepackage{amsthm}
\usepackage{enumerate}
\usepackage{bbm}
\usepackage{hyperref}
\usepackage{braket}
\usepackage{booktabs}
\usepackage{mathtools}
\usepackage{titlesec}
\usepackage[a4paper,top=2cm,bottom=2.5cm,left=2cm,right=2cm]{geometry}

\newcommand*{\Tr}{\mathrm{Tr}}

\newcommand*{\diff}{\text{d}}
\newcommand*{\coloneqq}{\mathrel{\vcenter{\baselineskip0.5ex \lineskiplimit0pt \hbox{\scriptsize.}\hbox{\scriptsize.}}} =}

\newcommand*{\h}{\mathcal{H}}

\newcommand*{\kbc}[3]{ |#1\rangle_{#3}\langle#2|}

\newtheorem{theorem}{Theorem}

\newtheorem{definition}[theorem]{Definition}

\makeatletter
\def\thmhead@plain#1#2#3{%
	\thmname{#1}\thmnumber{\@ifnotempty{#1}{ }\@upn{#2}}%
	\thmnote{{ \the\thm@notefont#3}}}
\let\thmhead\thmhead@plain
\makeatother

\newcommand{\virg}[1]{``#1''}